\documentclass[aps,prx,twocolumn,footinbib,showpacs,superscriptaddress,nobalancelastpage,floatfix,10pt]{revtex4-2}

\usepackage[english]{babel} 
\usepackage[utf8]{inputenc} 
\usepackage[T1]{fontenc} 
\usepackage{physics} 
\usepackage{amsmath} 
\usepackage{amssymb} 
\usepackage{amstext} 
\usepackage{graphicx} 
\usepackage[hidelinks]{hyperref} 
\usepackage[all]{hypcap} 
\usepackage[dvipsnames]{xcolor} 
\usepackage{lipsum} 
\usepackage{array} 
\usepackage[normalem]{ulem} 

\hypersetup{
    colorlinks=true,
    breaklinks=true,
    urlcolor=black,
    linkcolor=NavyBlue,
    citecolor=NavyBlue,
    bookmarksopen=true,
    pdftoolbar=false,
    pdfmenubar=false,
}

\newcommand{\ie}{i.e.}
\newcommand{\eg}{e.g.}

\newcommand{\e}{\textrm{e}} 
\newcommand{\hc}{\text{h.c.}} 
\renewcommand{\a}{\vb*{a}} 
\newcommand{\adag}{\vb*{a}^\dagger} 
\newcommand{\ta}{\tilde{\vb*{a}}} 
\newcommand{\tadag}{\tilde{\vb*{a}}^\dagger} 
\renewcommand{\b}{\vb*{b}} 
\newcommand{\bdag}{\vb*{b}^\dagger} 
\renewcommand{\r}{\vb*{r}} 
\newcommand{\rdag}{\vb*{r}^\dagger} 
\newcommand{\x}{\vb*{x}}
\newcommand{\p}{\vb*{p}}

\renewcommand{\sp}{\vb*{\sigma}_+} 
\newcommand{\sm}{\vb*{\sigma}_-} 
\newcommand{\sz}{\vb*{\sigma}_z} 
\newcommand{\sx}{\vb*{\sigma}_x} 
\renewcommand{\H}{\vb*{H}} 
\renewcommand{\L}{\vb*{L}} 
\newcommand{\vrho}{\vb*{\rho}} 
\DeclareMathOperator{\sign}{sign}

\newcolumntype{L}[1]{>{\raggedright\let\newline\\\arraybackslash\hspace{0pt}}p{#1}}
\newcolumntype{C}[1]{>{\centering\let\newline\\\arraybackslash\hspace{0pt}}p{#1}}
\newcommand\bigfrac[2]{\frac{\displaystyle #1}{\displaystyle #2}}

\begin{document}

\title{
Designing High-Fidelity Zeno Gates for Dissipative Cat Qubits
}

\author{Ronan Gautier}
\email[Corresponding author: ]{ronan.gautier@inria.fr}
\affiliation{Laboratoire de Physique de l'\'Ecole Normale Sup\'erieure, Inria, ENS, Mines ParisTech, Universit\'e PSL, Sorbonne Universit\'e, Paris, France}

\author{Mazyar Mirrahimi}
\affiliation{Laboratoire de Physique de l'\'Ecole Normale Sup\'erieure, Inria, ENS, Mines ParisTech, Universit\'e PSL, Sorbonne Universit\'e, Paris, France}

\author{Alain Sarlette}
\affiliation{Laboratoire de Physique de l'\'Ecole Normale Sup\'erieure, Inria, ENS, Mines ParisTech, Universit\'e PSL, Sorbonne Universit\'e, Paris, France}
\affiliation{Department of Electronics and Information Systems, Ghent University, Belgium}

\date{\today}

\begin{abstract}
Bosonic cat qubits stabilized with a driven two-photon dissipation are systems with exponentially biased noise, opening the door to low-overhead, fault-tolerant and universal quantum computing. However, current gate proposals for such qubits induce substantial noise of the unprotected type, whose poor scaling with the relevant experimental parameters limits their practical use. In this work, we provide a new perspective on dissipative cat qubits by reconsidering the reservoir mode used to engineer the tailored two-photon dissipation, and show how it can be leveraged to mitigate gate-induced errors. Doing so, we introduce four new designs of high-fidelity and bias-preserving cat qubit gates, and compare them to the prevalent gate methods. These four designs should give a broad overview of gate engineering for dissipative systems with different and complementary ideas. In particular, we propose both already achievable low-error gate designs and longer-term implementations.
\end{abstract}

\maketitle


\section{Introduction}\label{sec:intro}

The promise of quantum computing relies on the unpleasant predicament that a quantum system should be freely controllable but also very long-lived, two often conflicting requirements. To overcome this issue, a promising path towards fault-tolerance is quantum error correction~\cite{shor1995scheme,shor1996fault,cory1998experimental,kempe2001theory,bacon2006operator} with discrete-variable qubits. By redundantly encoding information, the eventual computational errors induced by environment noise can be resolved and corrected~\cite{fowler2012surface,chuang1995quantum,ai2021exponential,bonilla2021xzzx,acharya2022suppressing}. While the control of most discrete-variable qubits is now well-established~\cite{nakamura1999coherent,vion2002manipulating,koch2007charge,wallraff2005approaching,blais2021circuit,krinner2022realizing}, they are often error-prone by construction. For this reason, continuous-variable codes~\cite{lloyd1997analog,gottesman2001encoding,michael2016new,terhal2020towards,campagne2020quantum,albert2019pair,vuillot2019quantum} are attracting a rising interest thanks to their inherent robustness to errors and promise of low-overhead experimental setups~\cite{darmawan2021practical,noh2022low,schlegel2022quantum,hillmann2022quantum,kwon2022autonomous}.

In particular, bosonic cat qubits are encoded in opposite phase coherent states of a quantum harmonic oscillator~\cite{cochrane1999macroscopically,mirrahimi2014dynamically,puri2017engineering,puri2019stabilized,putterman2021colored,ruiz2022twophoton}. When stabilized with a driven two-photon dissipation, the so-called \textit{bit-flip} error consisting of flipping one computational state to the other is exponentially suppressed with the mean-number of photons in each coherent state. This can be viewed as a form of autonomous error correction, in which detection and correction of bit-flip errors is performed by a tailored interaction with the environment~\cite{leghtas2015confining,guillaud2019repetition,chamberland2020building,guillaud2021error,guillaud2022quantum,zapletal2022stabilization,gravina2022critical}. Exponentially large coherent state lifetimes have thus been demonstrated for dissipative cat qubits~\cite{lescanne2020exponential,berdou2022one}. 

The next key ingredient towards a hardware-efficient and fault-tolerant quantum processor based on cat qubits is to demonstrate single- and two-qubit physical gates that preserve the exponential protection against bit-flips~\cite{touzard2018coherent,puri2020bias,grimm2020stabilization,mizuno2022effect,regent2022high}. Besides this bit-flip suppression, these gates should admit \textit{phase-flip} errors well below the threshold of a discrete-variable error correcting code used for phase-flip correction, such as the repetition code. In this aim, several proposals have been put forward during the last two years to improve the fidelity of cat-qubit gates~\cite{xu2021engineering,gautier2022combined,xu2022autonomous}, with respective benefits and feasibility strongly depending on the particular experimental setup envisioned. 

The present work pursues with these research efforts, proposing alternatives whose performance is competitive at least in some contexts, and which may sometimes be naturally combined with these other strategies. We thus introduce four new designs for $Z(\theta)$, CNOT and Toffoli gates on dissipatively stabilized cat qubits to help mitigate the incoherent phase errors induced by those gates. Indeed, while the prevalent method of gate engineering based on the Zeno effect~\cite{mirrahimi2014dynamically,guillaud2019repetition} has the great benefit of simplicity, it also features relatively large gate errors and a poor scaling with the relevant experimental parameters that may limit its practical use~\cite{chamberland2020building}.

To answer this limitation, our main approach is to reduce the logical information which the two-photon dissipation carries away to the environment under Zeno driving and which induces phase-flip backaction. Our first two designs maintain the same Zeno drives and interaction with the buffer mode mediating two-photon dissipation, but they feed information back from the buffer state to the cat-qubit system before it can leak out to the environment. With this principle, we were able to improve gate fidelities by up to two orders of magnitude with realistic experimental parameters. Our other two designs instead modify the Zeno gate drive, to avoid pushing any information from the cat qubit to the dissipative buffer in the first place. One solution, based on locally flat Hamiltonians, promises a polynomial improvement in the scaling of gate errors with the cat size. Our final design demonstrates exponentially small $Z(\theta)$ gate errors with a tailored dissipation to an ancillary qubit. While its generalization to CNOT gates is beyond the current state of the art, it provides a new way to leverage dissipation for gate engineering.

The paper is organized as follows. We begin in Section~\ref{sec:catreview} by reviewing dissipative cat qubits and their standard gates implementation. In Section~\ref{sec:gateerrors}, we provide a new perspective on the origin of gate errors using the buffer mode. Section~\ref{sec:summary} gives a short summary of the different gate error mitigation designs introduced in this paper. And finally, each design is detailed separately in Sections \ref{sec:photodetection} -- \ref{sec:discrete}. We conclude in Section~\ref{sec:conclusions}.


\section{Review of Cat Qubit Gates}\label{sec:catreview}
\subsection{Stabilization and encoding}
In the dissipative stabilization of cat qubits, a harmonic oscillator exchanges pairs of photons with its environment both through a driven dissipation process and a two-photon pump. The Lindblad master equation that governs this oscillator is
\begin{equation}\label{eq:twophdiss}
    \frac{d \vrho}{dt} = \kappa_2 \mathcal{D}[\a^2 - \alpha^2] \vrho
\end{equation}
where $\mathcal{D}[\L]\vrho = \L \vrho \L^\dagger - \{ \L^\dagger \L, \vrho\}/2$ is the dissipation superoperator, $\a$ denotes the annihilation operator of the cat qubit mode, $\kappa_2$ the rate of two-photon dissipation, and $\alpha$ the cat qubit amplitude. Throughout this paper, we assume $\alpha \in \mathbb{R}$ unless specified otherwise. To engineer this unusual dissipation, an ancillary buffer mode --- not necessarily harmonic --- is often introduced to mediate the exchange of photon pairs between the oscillator and its environment, as depicted in Fig.~\hyperref[fig:bufferdynamics]{1(a)}. This results in a two-mode Lindblad master equation,
\begin{equation}~\label{eq:masteqbuffer}
    \frac{d \vrho}{dt} = -i \left[ \H_{AB}, \vrho \right] + \kappa_b \mathcal{D}[\b] \vrho
\end{equation}
where $\H_{AB} = g_2 (\a^2 - \alpha^2) \bdag + \hc$ is a driven two-to-one photon exchange Hamiltonian with $\b$ the annihilation operator of the ancillary buffer mode. In the limit of $\kappa_b \gg g_2$, this additional mode can be adiabatically eliminated~\cite{azouit2015convergence,azouit2016well,azouit2017towards,forni2018adiabatic} to retrieve the single-mode model of~\eqref{eq:twophdiss} with a typical two-photon loss rate $\kappa_2 \equiv 4 g_2^2 / \kappa_b$. Both systems feature a degenerate subspace of steady states of dimension 2 in which the cat qubit is encoded. In the single-mode system, the associated code words of this qubit are defined as
\begin{equation}\label{eq:logicalstates}
    \begin{split}
        \ket{0_L} &\equiv \frac{1}{\sqrt{2}} \left( \ket{\mathcal{C}_\alpha^+} + \ket{\mathcal{C}_\alpha^-}\right) = \ket{\alpha} + \mathcal{O}(\e^{-2|\alpha|^2}) \\
        \ket{1_L} &\equiv \frac{1}{\sqrt{2}} \left( \ket{\mathcal{C}_\alpha^+} - \ket{\mathcal{C}_\alpha^-}\right) = \ket{-\alpha} + \mathcal{O}(\e^{-2|\alpha|^2})
    \end{split}
\end{equation}
where $\ket{\pm \alpha}$ are opposite-phase coherent states and $\ket{\mathcal{C}_\alpha^\pm} \equiv \mathcal{N}_\pm \left(\ket{\alpha} \pm \ket{-\alpha} \right)$ are Schr\"{o}dinger cat states of even and odd photon-number parity.

This dissipative stabilization of cat qubits is quite remarkable for its biased robustness to errors that are local in phase space, whether they may come from spurious Hamiltonians --- for instance neglected in some underlying rotating wave approximation --- or undesired couplings of the oscillator to its environment. Indeed, thanks to the coherent states being localized in opposite sides of phase space, it is expected that the bit-flip error rate $\Gamma_X$ of a dissipative cat qubit is suppressed exponentially in the cat mean-number of photons $|\alpha|^2$, according to $\Gamma_X = \Gamma_{X,0} \exp(-2 |\alpha|^2)$~\cite{mirrahimi2014dynamically, lescanne2020exponential}. Although difficult to compute, the prefactor $\Gamma_{X,0}$ typically depends on terms that cause leakage out of the qubit codespace and may vary during the operation of the device. On the contrary, the qubit phase information is encoded in the photon-number parity. This degree of freedom is not protected by the stabilization mechanism, and it typically becomes linearly more fragile as the cat mean-number of photons $|\alpha|^2$ increases. Hence, the idea behind cat qubits is to benefit from the exponential hardware protection against bit-flips, and to concentrate error-correction efforts on phase-flips with hopefully low-enough error rates.

\subsection{Bias-preserving gates} 

For this paradigm to work, the noise bias must be preserved when operating gates on the physical qubits, meaning that we should never convert phase errors into bit errors (or introduce significant bit errors in any other ways). In~\cite{guillaud2019repetition}, a set of bias-preserving physical operations is proposed for universal quantum computation with cat qubits. This set comprises Pauli $X$, $Z(\theta)$ rotations, CNOT and Toffoli gates, in addition to the preparation and measurement of cat states in the orthonormal basis $\ket{\pm_L} = \ket{\mathcal{C}_\alpha^\pm}$. Such a gate set allows for forward concatenation with various logical error-correcting codes such as the repetition code~\cite{guillaud2019repetition}, the XZZX code~\cite{bonilla2021xzzx} or a rectangular surface code~\cite{chamberland2020building}. In the remainder of this section, we highlight some physical gate design elements that will be useful for this paper; a more complete review can be found in~\cite{guillaud2019repetition}.

\subsubsection{Pauli \texorpdfstring{$X$}{X}}

The Pauli $X$ gate corresponds to a $\pi$-phase delay on the oscillator, or equivalently, to a cat qubit codespace rotation in phase space for some time $T = \pi / \Delta$, where $2\Delta$ is the buffer frequency detuning with respect to the cat-qubit frame. This exchanges the two computational basis states. From a complementary viewpoint, the engineered dynamics should stabilize the cat qubit codespace with~\eqref{eq:twophdiss}, but in a frame that rotates at the detuned frequency. Working this back to cat-qubit frame, we see that we must engineer the dynamics of
\begin{equation}\label{eq:fullXterms}
    \frac{d \vrho}{dt} = - i [\Delta \adag \a, \vrho] + \kappa_2 \mathcal{D}[\a^2 - \alpha^2\e^{-2i\Delta t}] \vrho.
\end{equation}
Since this master equation preserves photon parity, no phase errors are induced by this gate design. 

While the implementation of both terms of~\eqref{eq:fullXterms} is well-established, one can in principle choose to engineer only one of the two terms. If only the second term of~\eqref{eq:fullXterms} is kept, the cat state is pulled by the time-varying setpoint of the dissipation, incurring negligible leakage and bit-flips for a sufficiently slow codespace rotation. With the Hamiltonian alone, the rotation is performed exactly, but the stabilizing effect of two-photon dissipation is turned off during the gate. For fast-enough gates however, the design is still bias-preserving under local errors, and codespace leakage can be suppressed after the gate once stabilization is turned back on.

\subsubsection{\texorpdfstring{$Z(\theta)$}{Z} rotations}

Cat-qubit $Z(\theta)$ rotations require the accumulation of a different phase on the $\ket{\alpha}$ and $\ket{-\alpha}$ components of the codespace. Since this exact evolution is not directly accessible with simple experimental means, the standard proposal is an approximate one based on the Zeno effect. The combination of a small drive displacing the state in phase-space and of the dominating dissipation of~\eqref{eq:twophdiss} pulling the state back to the codespace induces to first order an effective phase-shift inside a slightly deformed codespace, and hence the desired gate~\cite{mirrahimi2014dynamically}. The master equation to be engineered takes the standard form,
\begin{equation}
    \frac{d \vrho}{dt} = - i [\H_Z, \vrho] + \kappa_2 \mathcal{D}[\a^2 - \alpha^2] \vrho
\end{equation}
with, for real-valued $\alpha$, a drive Hamiltonian
\begin{equation}\label{eq:zdrive}
    \H_Z \equiv \varepsilon_Z (\adag + \a)
\end{equation}
for which the gate angle reads $\theta = 4\alpha \int \varepsilon_Z dt$. To second order, the Zeno effect involves phase decoherence that scales as $(\varepsilon_Z / \kappa_2)^2$, as discussed more extensively in Section~\ref{sec:gateerrors}. 

\subsubsection{CNOT and Toffoli}

By definition, a CNOT gate is a Pauli $X$ gate on a target qubit conditioned on the state of a control qubit along its $Z$ axis, or equivalently, a Pauli $Z$ gate on a control qubit conditioned on the state of a target qubit along its $X$ axis. Therefore, the standard design of CNOT or Toffoli gates for cat qubits involves a combination of the $X$ and $Z$ gate implementations, and of their respective issues. From the viewpoint of the $X$ gate, the dynamics of~\eqref{eq:fullXterms} should be applied conditionally on the computational state of a control qubit. As introduced in~\cite{guillaud2019repetition}, this can be done with the two-mode master equation
\begin{equation}
    \frac{d \vrho}{dt} = -i [\H_{CX}, \vrho] + \kappa_2 \mathcal{D}[\L_C] \vrho + \kappa_2 \mathcal{D}[\L_T(t)]\vrho
\end{equation}
where $\L_C \equiv \a_C^2 - \alpha^2$ is the standard two-photon dissipation of~\eqref{eq:twophdiss} on the control qubit, $\L_T(t)$ is a time-dependent dissipation on the target qubit,
\begin{equation}\label{eq:cnotdiss}
    \L_T(t) \equiv \a_T^2 - \frac{\alpha}{2}(\a_C + \alpha) + \frac{\alpha}{2}(\a_C - \alpha) \e^{-2i\Delta t}
\end{equation}
and the coupling Hamiltonian reads
\begin{equation}\label{eq:cnotff}
    \H_{CX} \equiv \varepsilon_{CX} (\adag_C + \a_C - 2\alpha) (\adag_T \a_T - n_p) .
\end{equation}
Here $\a_{C/T}$ denote the control and target qubit modes respectively, $n_p$ is any even integer close to $|\alpha|^2$, and $\varepsilon_{CX}(t) = \Delta(t) / 4\alpha$. Adding a second control mode in~\eqref{eq:cnotdiss} and~\eqref{eq:cnotff} yields the Toffoli gate operators. Similarly to the $X$ gate, Hamiltonian~\eqref{eq:cnotff} may be dropped to ease experimental requirements as long as the setpoint of the dissipation in~\eqref{eq:cnotdiss} rotates slowly enough to adiabatically pull the state. Alternatively, the dissipation of~\eqref{eq:cnotdiss} can also be dropped, in which case the Hamiltonian~\eqref{eq:cnotff} alone induces the intended target qubit rotation, but at the cost of turning off target qubit stabilization temporarily during the gate.

The viewpoint of the $Z$ gate clarifies the impact of conditioning on the control qubit. The link is most direct when the dissipator~\eqref{eq:cnotdiss} is dropped, as proposed in~\cite{gautier2022combined}. Indeed, Hamiltonian~\eqref{eq:cnotff} together with two-photon dissipation~\eqref{eq:twophdiss} on the control qubit, amounts to the same Zeno dynamics as for the $Z(\theta)$ rotation; the only difference being that the drive amplitude $\varepsilon_Z$ in~\eqref{eq:zdrive} is now conditioned on the photon number in the target qubit state. This achieves the required gate since, at first order, the control qubit undergoes an even number of $Z(\pi)$ gates for all even Fock states of the target qubit, and respectively an odd number of $Z(\pi)$ gates for all odd Fock states, hence refocusing on the even and odd cat codespaces at the end of the CNOT gate. At second order though, we see that like for the $Z$ gate, the process induces phase decoherence on the control cat-qubit.

\subsubsection{Gate errors}

During each of these gates, there are two main effects that induce phase errors. The first is spontaneous emission of photons from the oscillator, with dissipation operator $\mathcal{D}[\a]$. This causes phase errors linearly in time and in $|\alpha|^2$ independently of the ongoing gate, and also results in correlated errors for multi-qubit gates~\cite{chamberland2020building}. Since this source of decoherence can only be mitigated by increasing oscillator lifetime or by changing the overall qubit encoding, it will not be considered any further in this paper. However, it could in principle feature non-trivial dynamics once combined with gate proposals, so a complete model of errors is studied in Appendix~\ref{sec-apdx:errormodel}. In addition, single-photon losses motivate the need for faster gate operations (thus increasing overall fidelities) and limits the benefits granted by increasing $|\alpha|^2$. 

Phase errors are also induced directly by gate processes, but only on modes for which parity-switching dynamics is engineered. In particular, target qubits of multi-qubit gates do not suffer from such errors, and hence most of the dynamics of interest occurs on control qubits. Consequently, $Z(\theta)$ rotation gates provide much of the important physics, and strategies to mitigate cat qubit gate errors can all be understood within the scope of this single-qubit gate. This paper is therefore devoted to the mitigation of gate-induced errors, and often begins with the design of $Z(\theta)$ gates before generalizing.


\section{Zeno Gate Errors}\label{sec:gateerrors}

To illustrate the origin of gate-induced errors and how they can be mitigated by tampering with the ancillary buffer mode, let us consider the full Hamiltonian engineered for $Z(\theta)$ gates in the presence of the buffer mode,
\begin{equation}~\label{eq:zgate-buffer}
    \H = \H_{AB} + \H_Z.
\end{equation}
Together with the high damping rate of the buffer, the first term mediates the two-photon dissipation while the second term drives the required gate. Let us move into the shifted Fock basis as introduced in~\cite{chamberland2020building} (see also Appendix~\ref{sec-apdx:gatereview} for a short review). This change of basis reads
\begin{equation}\label{eq:sfb-transf}
    \a \rightarrow \sz \otimes (\ta + \alpha)
\end{equation}
and effectively represents a displaced oscillator where the displacement is conditional on $\sz$, the Pauli operator corresponding to the cat qubit logical state. This change of basis is non-orthonormal and approaches degeneracy at high Fock states, but it is sufficiently close to a regular change of coordinates for states close to $\ket{\pm \alpha}$, and thus adequate for our investigation of local errors in the short time limit. Then, $\ta$ is a gauge mode that models a local oscillator around the $\ket{\pm \alpha}$ coherent states; in particular, $\ta$ in vacuum is equivalent to being perfectly inside the cat-qubit encoding space. With this definition, the Hamiltonian~\eqref{eq:zgate-buffer} reads ($\alpha \in \mathbb{R}$)
\begin{equation}
    \vb*{\tilde{H}} = g_2 (\ta^2 + 2\alpha \ta) \bdag + \varepsilon_Z \sz (\tadag + \alpha) + \hc
\end{equation}
The ideal Hamiltonian that implements the $Z(\theta)$ rotation of the qubit (and nothing else) now appears as the term $(\alpha \varepsilon_Z\sz + \hc)$. In the following, we move into the rotating frame of this Pauli Hamiltonian to simplify the analysis. In addition, in the limit of a small Zeno drive, the effective displacements on the gauge mode $\ta$ and on the buffer mode $\b$ are small, such that we can neglect the second-order term $\ta^2 \b^\dagger + \hc$. The corresponding Hamiltonian thus reads
\begin{equation}\label{eq:zgate-sfb}
    \vb*{\tilde{H}}' \approx 2\alpha g_2 \ta \bdag + \varepsilon_Z \sz \tadag + \hc
\end{equation}
Writing the master equation with the Hamiltonian of~\eqref{eq:zgate-sfb} and in the Heisenberg picture~\cite{breuer2002theory} for both $\ta$ and $\b$ yields a set of coupled equations,
\begin{subequations}
    \begin{align}
        &\dot{\ta} = -2i \alpha g_2 \b - i \varepsilon_Z \sz \\
        &\dot{\b} = -2i \alpha g_2 \ta - \kappa_b \b / 2 \label{eq:bufferdyn-coupled}
    \end{align}
\end{subequations}
Decoupling these equations gives the second-order differential equations
\begin{subequations}
    \begin{align}
        &\ddot{\ta} + \tfrac{1}{2}\kappa_b \dot{\ta} + \nu^2 \ta = - \tfrac{i}{2} \kappa_b \varepsilon_Z \sz - i \dot{\varepsilon}_Z \sz \\
        &\ddot{\b} + \tfrac{1}{2}\kappa_b \dot{\b} + \nu^2 \b = -\nu \varepsilon_Z \sz
    \end{align}
\end{subequations}
where $\nu \equiv 2 \alpha g_2$, and we have used that $d \sz / dt = 0$ since $\sz$ commutes with~\eqref{eq:zgate-sfb}. Finally, making the inverse change of basis on the cat qubit mode, \ie~$\sz \otimes (\ta + \alpha) \rightarrow \a$ yields
\begin{subequations}\label{eq:catbufferdyn}
    \begin{align}
        &\ddot{\a} + \tfrac{1}{2} \kappa_b \dot{\a} + \nu^2 \a = \nu^2 \alpha \sz - \tfrac{i}{2} \kappa_b \varepsilon_Z - i \dot{\varepsilon}_Z \\
        &\ddot{\b} + \tfrac{1}{2} \kappa_b \dot{\b} + \nu^2 \b = -\nu \varepsilon_Z \sz \label{eq:bufferdyn}
    \end{align}
\end{subequations}
Both the cat qubit and buffer mode are thus described by a damped harmonic oscillator equation with natural frequency $\nu$ and damping rate $\kappa_b / 2$. These oscillators are further driven out of equilibrium by the $\varepsilon_Z$ drive. 

\begin{figure}[!t]
         \centering
         \includegraphics[width = \columnwidth]{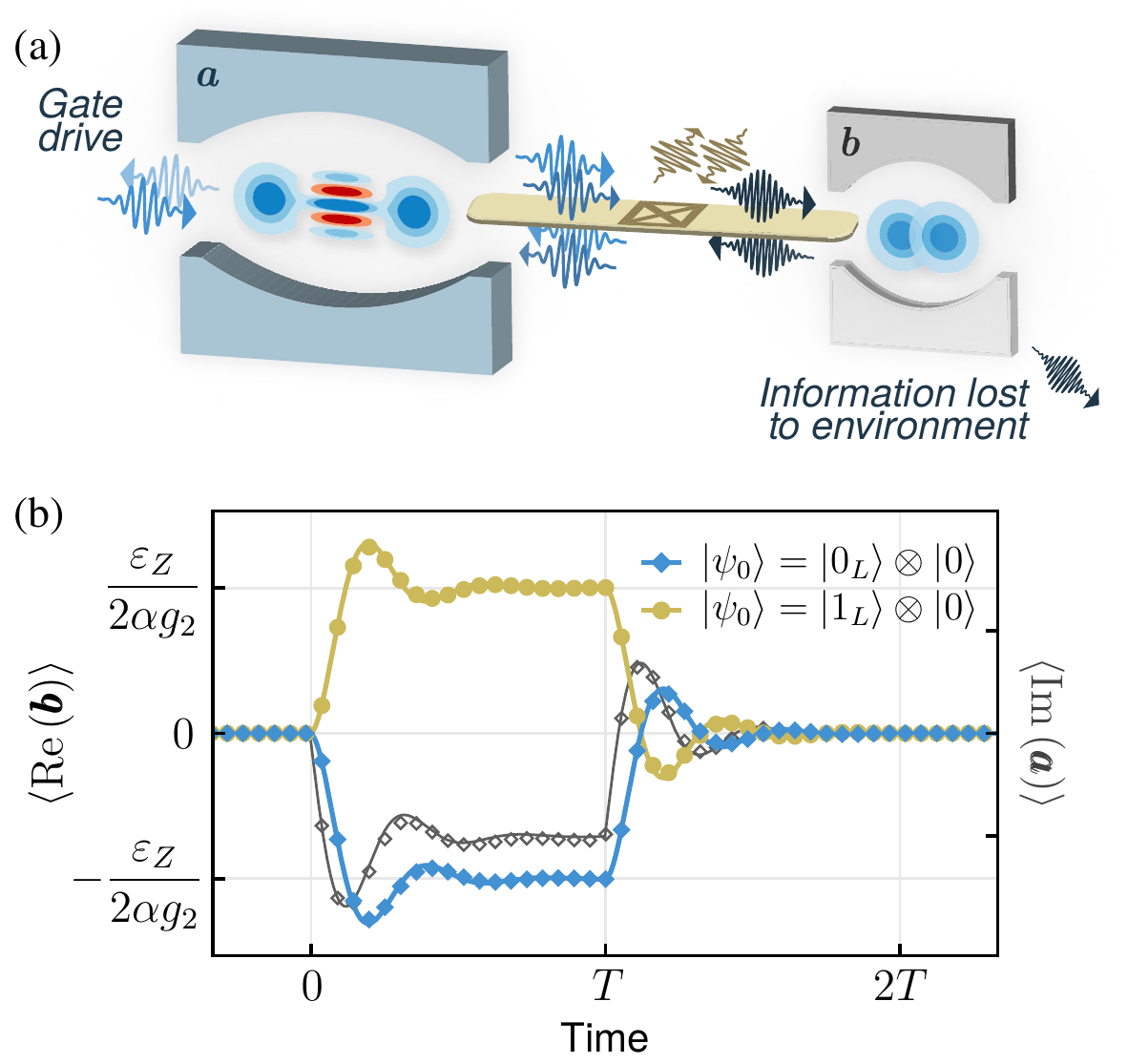}
\label{fig:bufferdynamics}
\vspace{-0.5cm}
\caption{
    (a) Cat qubit mode $a$ stabilized by a two-to-one photon exchange with buffer mode $b$, mediated by a nonlinear circuit element (middle).
    The gate drive, by displacing the cat mode, induces a small splitting of the buffer mode conditioned on the cat state. Information is then lost to the environment through the high damping rate of the buffer, hence inducing incoherent gate errors. (b) Average displacement of the buffer (color, left scale) and cat mode (black, right scale) before, during and after a Zeno $Z(\pi)$ gate of duration $T$, with a cat mode initialized in $\ket{0_L/1_L} \approx \ket{\pm\alpha}$ and buffer initialized in vacuum. Markers show numerical data obtained from integration of the full master equation. Lines show Eqs.~\eqref{eq:catbufferdyn}.
}
 \end{figure}

On the cat qubit mode, we naturally find that equilibrium is given by $\a_{eq} = \alpha \sz$ when $\varepsilon_Z = 0$, which corresponds to the computational states of cat qubits up to exponentially small corrections that were neglected by introducing the shifted Fock basis. We also find that, for $\varepsilon_Z > 0$, the mode is displaced along the $\langle \text{Im}(\a) \rangle$ quadrature independently of $\sz$, as expected. 

On the buffer mode, it is quite interesting to note that the right-hand side term is proportional to $\sz$, such that the buffer mode is displaced in opposite directions depending on the computational state of the cat qubit mode. In other words, the equilibrium position for the buffer mode is $\b_{eq} = - \varepsilon_Z / \nu \sz$ and more importantly, $\b \propto \sz$ at all times. Because the buffer mode is largely damped, the environment obtains information about the state of the cat qubit mode through this effect. As a consequence, measurement of the bit value by the environment dissolves the information contained in the superposition of these bit states, and thus induces phase errors. In fact, it is simple to re-derive the well-known result of the $Z(\theta)$ gate non-adiabatic error rate by replacing $\b$ with $\b_{eq}$ in the dissipator of~\eqref{eq:masteqbuffer} and integrating over the gate duration. This yields
\begin{equation}\label{eq:pz0}
    p_Z = p_Z^{(0)} \equiv \frac{\theta^2}{16 |\alpha|^4 T} \frac{\kappa_b}{4 g_2^2}
\end{equation}
which is the same result as in~\cite{chamberland2020building} with $\kappa_2 \equiv 4 g_2^2 / \kappa_b$.

Figure~\hyperref[fig:bufferdynamics]{1(b)} shows the average displacement of the buffer mode along the $\langle \text{Re}(\b) \rangle$ quadrature before, during, and after a $Z(\pi)$ gate of duration $T$. Depending on the initial state of the cat qubit mode, either $\ket{0_L}$ (blue) or $\ket{1_L}$ (yellow), the buffer mode is displaced in one direction or the other. In fact, we observe the dynamics of a damped oscillator with equilibrium position $\b_{eq} = \pm \varepsilon_Z / \nu$ for $0 < t < T$ and $ \b_{eq} = 0$ for $t > T$, in excellent agreement with our analysis based on the approximate model~\eqref{eq:catbufferdyn}. The displacement of the cat qubit mode is also shown in black and is independent of the initial state, as expected.

This derivation is meant to provide intuition to the reader about the origin of gate errors. With this intuition in mind, the following sections will introduce multiple gate designs to reduce the errors induced by cat qubit Zeno dynamics, beginning with a summary.


\section{Summary of Gate Designs}\label{sec:summary}

\begin{table*}[t]
    \caption{Comparison of different designs of $Z(\theta)$ gates for dissipative cat qubits. All designs but the last one can be generalized to two- and three-qubit CNOT and Toffoli gates. For the second and last designs, $\sp$ denotes a Pauli creation operator on some ancillary qubit. $\H_{AB}$ and $\H_Z$ are defined in~\eqref{eq:masteqbuffer} and~\eqref{eq:zdrive} respectively, and $\a_\theta = \sin(\theta/2) \a + i \cos(\theta/2) \alpha$. In the last column, $p_Z$ denotes the probability of gate errors over a single $Z(\pi)$ gate of duration $T$.}
    \renewcommand{\arraystretch}{1.8}
    \begin{tabular}{L{0.095\textwidth} L{0.20\textwidth} L{0.27\textwidth} L{0.20\textwidth} L{0.195\textwidth}}
        \hline \hline
         & & Hamiltonian $\H$ & Dissipator $\mathcal{D}$ & Gate Errors \\[0.4ex]
        
        \hline
        Ref.~\cite{mirrahimi2014dynamically, chamberland2020building} & Standard Zeno & $\H_{AB} + \H_Z$ & $\kappa_b\mathcal{D}[\b]$ & $p_Z^{(0)} \equiv \bigfrac{\pi^2}{16 |\alpha|^4 T} \bigfrac{\kappa_b}{4 g_2^2}$ \\
    
        Ref. \cite{gautier2022combined} & Combined dissipation and TPE Hamiltonian & $\H_{AB} + \H_Z + \H_{TPE}$ with \newline $\H_{TPE} \equiv g'_2(\a^2 - \alpha^2) \sp + \hc$ & $\kappa_b\mathcal{D}[\b]$ & $p_Z = \bigfrac{1}{1+ (2 g'_2 / \kappa_2)^2} p_Z^{(0)}$ \\

        Sec. \ref{sec:photodetection} & Buffer photodetection with classical feedback & $\H_{AB} + \H_Z$ & $\kappa_b\mathcal{D}[\b]$ \newline (photodetected) & $p_Z \gtrsim (1-\eta) p_Z^{(0)}$ \newline (detection efficiency $\eta$)\\

        Sec. \ref{sec:autonfeedback} & Cat-buffer autonomous feedback & $\H_{AB} + \H_Z$ & $\kappa_{ab} \mathcal{D}[\a\b]$ & $p_Z = \mu\, p_Z^{(0)}$ \newline with $ \mu \gtrsim 0.02$\\

        Sec. \ref{sec:flathamil} & Locally flat Hamiltonian & $\H_{AB} + \H_{Z, N}$ with \newline  $\H_{Z, N} = \varepsilon_Z \sum_{n=0}^N c_n (\a+\adag)^{2n+1}$ & $\kappa_b\mathcal{D}[\b]$ & $p_Z = \nu |\alpha|^{-2N} p_Z^{(0)}$ \newline with $ \nu \sim 1$\\
        
        Sec. \ref{sec:discrete} & Discrete jump & $\H_{AB}$ & $\kappa_b\mathcal{D}[\b] + \kappa_Z \mathcal{D}[\a_\theta \sp]$ & $p_Z = \exp(-\kappa_Z |\alpha|^2 T)$ \\[0.3em]
        \hline \hline
    \end{tabular}
    \label{tab:summary}
\end{table*}

With the analysis of the previous section in mind, we understand that the loss of phase information during gates is due to the conditional displacement of the buffer mode which is then measured by the environment. This is indeed the only non-unitary channel, hence how quantum information can be lost. Compared to a system with no ancillary buffer in which the information would be directly lost to the environment, the delay provided by the buffer mode can be exploited to reduce gate-induced phase errors without tampering with the ongoing gate.

In this paper, we introduce two methods that rely on kicking back qubit information that has been transferred into the buffer mode, and two methods that rely on the reduction of information transfer to the buffer mode in the first place. They are all summarized in Table~\ref{tab:summary} along with the regular Zeno-based gate of~\cite{mirrahimi2014dynamically} and the combined confinement method which we have recently proposed in~\cite{gautier2022combined}. Although the table only tackles these methods in the scope of the $Z(\theta)$ gate, they can all be generalized to multi-qubit gates and aim to represent a wide range of ideas for the mitigation of dissipative cat qubit gate errors. Furthermore, some of these ideas can in principle be combined to attain even higher fidelities.

\subsection{Buffer photodetection with classical feedback}
In the first method, detailed in Section~\ref{sec:photodetection}, the principle is to retrieve the information leaking out by directly measuring the field coming out of the buffer mode, instead of letting it get lost to the environment. An appropriate feedback action can then restore this information back into the cat qubit system. The measurement is a photon counter. The feedback action corresponds to additional Pauli $Z$ gates performed either in software, or through a modification of the gate drive amplitude and/or duration. Alternatively, one can pursue a heralded gate. An immediate limitation of this technique is detector efficiency which will directly limit the proportion of information loss which we can counter with respect to~\eqref{eq:pz0}. Despite this limitation, this method could become viable with the rapid improvement in circuit-integrated photodetectors, and is in any case instructive for the following design.

\subsection{Cat-buffer autonomous feedback}
A second method based on the buffer information is presented in Section~\ref{sec:autonfeedback}. The idea is again that feedback on the cat qubit photon-number parity is applied after detection of a buffer mode photon. However, instead of actually applying a measurement and action, this loop is now applied autonomously thanks to a tailored dissipation operator. This tailored dissipation takes the form $\mathcal{D}[\a\b]$ such that any time the buffer loses a photon to the environment --- and by doing so swaps the cat qubit parity, as we showed in Section~\ref{sec:gateerrors} ---, a second parity-switch is applied on the cat qubit through the loss of a single cavity photon. With this autonomous feedback, ``detection efficiency'' is in principle perfect and a parameter-independent improvement in fidelity of about two orders of magnitude is numerically demonstrated. This design can further be generalized to any multi-qubit C$^n$X gate with no additional multi-qubit interactions. The residual phase errors with this design are due to second-order effects that cause imperfections in the applied feedback. 

\subsection{Locally flat Hamiltonian}
This third method is the first of a second strategy which consists in minimizing the amount of computational information transferred from the memory to the buffer mode by the gate process. Doing so, the environment cannot in turn receive information by measuring buffer output photons, so qubit coherence is preserved. This gate design is presented in Section~\ref{sec:flathamil}. It introduces drive Hamiltonians of the form $\H = f(\x)$ where $\x = \a+\adag$ is the real field quadrature operator on the cat qubit mode, and such that $f(x)$ describes a quasi-potential that is locally flat around $x = \pm \alpha$ but with different mean values $f(\alpha) \neq f(-\alpha)$. The first condition ensures that the drive Hamiltonian is approximately constant over the $\x$-eigenstates spanned by each computational state $\ket{\pm \alpha}$, such that it induces almost no dynamics on them and they stay inside the codespace throughout the gate. The second condition ensures that each computational state picks up a different phase, hence a rotation about the $Z$ axis. Such Hamiltonians can be engineered with various orders of odd polynomials in $\x$, and, in the limit of a polynomial of infinite order, $f(x)$ would essentially become the sign function; then exponentially low gate errors in $|\alpha|^2$ are demonstrated, only limited by the finite overlap of coherent states $\ket{\alpha}$ and $\ket{-\alpha}$.

\subsection{Discrete jump}
Section~\ref{sec:discrete} introduces a cat qubit gate design based on a tailored `discrete' dissipation. Concretely, through interaction with an ancillary qubit mode, exactly a single photon is subtracted from the system and the cat state is mapped onto the same state with a gate applied, and so in a discrete manner. Furthermore, for the specific angle $\theta = \pi$, the system stays exactly within its codespace during gates, so no information is transmitted to the environment, opening the door to exponentially low gate errors.

The main limitation of this gate design lies in the introduced ancillary qubit. Indeed, ancillary qubit relaxation would result in additional $Z(\theta)$ rotations for $Z(\theta)$ gates, or additional $Z(\pi)$ rotations for $CZ$ gates. In the case of CNOT or Toffoli gates, it would induce $X$ gate errors, thus killing the error bias. While the dissipator to be engineered for the $Z(\theta)$ and CZ gates is feasible with current state of the art experiments, a viable way to engineer the required CNOT and Toffoli dissipators thus remains to be found.

\subsection{Combined dissipation and Two-Photon Exchange Hamiltonian}
Finally, the combined two-photon exchange Hamiltonian and two-photon dissipation method introduced in~\cite{gautier2022combined} can also yield high-fidelity gates. Indeed, thanks to the additional two-photon exchange Hamiltonian confinement, the effective displacement of the cat qubit mode during a gate is greatly reduced such that less information is transmitted to the buffer mode and thus to the environment. This provides a very simple design for gate error mitigation, especially considering the similarity between this Hamiltonian and the one required for the usual dissipative confinement. This design is not further treated in this paper, and we refer to~\cite{gautier2022combined} for more details.

\subsection{Combining designs}
As a final remark to this summary, let us note that the gate designs introduced here can be combined together to further improve gate fidelities. As an example, it could be highly favorable to engineer a dissipative cat qubit with the correlated dissipator in $\mathcal{D}[\a\b]$, together with a locally flat Hamiltonian to drive gates, plus possibly an optimized activation profile like in~\cite{xu2021engineering}.
We however treat each design separately for clarity.


\section{Buffer Photodetection with Classical Feedback}\label{sec:photodetection}
\subsection{Design principle} 

Consider the model of~\eqref{eq:masteqbuffer} with a photodetector measuring the output field of the buffer mode. It is governed by a stochastic master equation (SME) that reads~\cite{steck2007quantum}
\begin{equation}\label{eq:sme}
    \begin{split}
        d \vrho = &-i \left[\H_{AB} + \H_Z, \vrho\right] dt  \\
        &+ \kappa_b \mathcal{D}_\eta[\b] \vrho \, dt + \mathcal{J}[\b] \vrho \, dN_\eta
    \end{split}
\end{equation}
where $\H_{AB}$ denotes two-photon exchange between cat and buffer modes, $\H_Z$ the drive Hamiltonian,
\begin{equation}
    \mathcal{D}_\eta[\b] \vrho = \mathcal{D}[\b] \vrho - \eta (\b \vrho \bdag - \langle \bdag \b \rangle \vrho)
\end{equation}
is a corrected dissipation that accounts for the backaction of no-detection events, and
\begin{equation}\label{eq:smeops-jump}
    \mathcal{J}[\b] \vrho = \frac{\b \vrho \bdag}{\langle \bdag \b \rangle} - \vrho
\end{equation}
is a stochastic jump process that accounts for detection events. Here, $\eta \in [0,1]$ is the detector efficiency and $d N_\eta$ denotes a stochastic counting process such that it is unity with probability $\langle d N_\eta \rangle = \eta \kappa_b \langle \bdag \b \rangle dt$ and zero otherwise. Like the non-stochastic master equation, the SME of~\eqref{eq:sme} features the cat qubit codespace with the buffer in vacuum as its only subspace of steady states. If the system steers away from these steady states --- a process otherwise known as `codespace leakage' which is in particular induced during gates --- then the buffer mode will get populated through the two-photon exchange interaction and it will output photons to the detector through its large damping rate. During this process, the detector may click, depending on the average buffer mode population and on detection efficiency.

\begin{figure}[!t]
    \centering
    \includegraphics[width = \columnwidth]{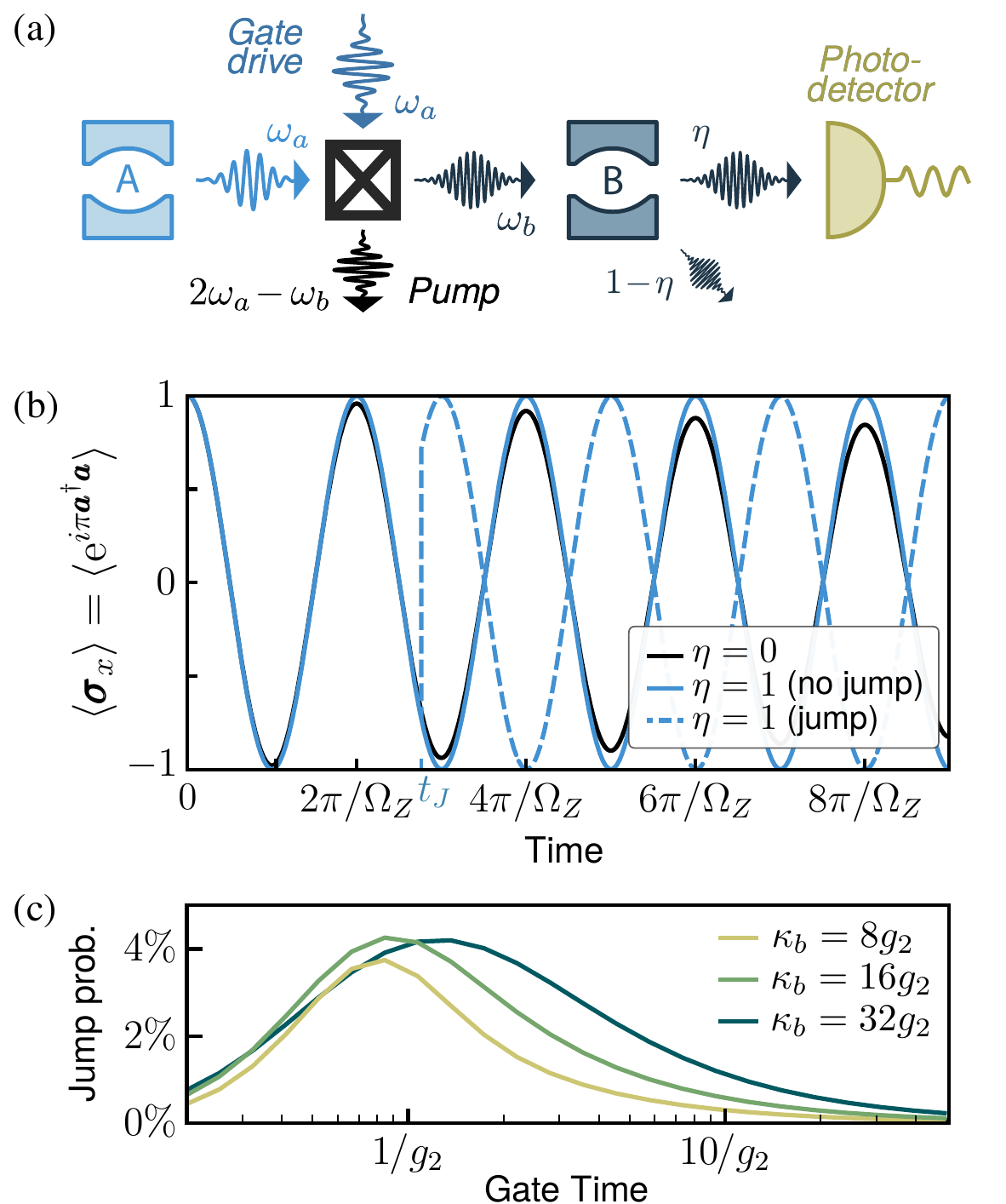}
    \vspace{-0.5cm}
    \label{fig:PD-rabi-osc}
    \caption{
        (a) Four-wave mixing process showing single cavity photons converted upwards into buffer photons when the gate drive is on. They are then measured by a photodetector on the buffer output. (b) Rabi oscillations with a photodetector on the buffer mode output. The system is initialized in $\ket{+}_L \otimes \ket{0}$ and a constant single-photon drive~\eqref{eq:zdrive} is turned on at $t=0$. Parity of the cat qubit mode is monitored against time for $\eta = 0$ (no detector) and $\eta=1$ (perfect detector) both for a no-jump trajectory and a single-jump trajectory. In this numerical simulation, $\Omega_Z \equiv 4\alpha \varepsilon_Z = \pi / g_2$, $\kappa_b = 8 g_2$ and $|\alpha|^2 = 8$. 
        (c) Total probability of at least one jump to occur during a $Z(\pi)$ gate, for $|\alpha|^2 = 8$.
    }
\end{figure}

An alternate explanation to this process is found from the viewpoint of four-wave mixing, as represented in Fig~\hyperref[fig:PD-rabi-osc]{2(a)}. Indeed, during the dynamics, it is possible for 1 photon of the gate drive and 1 photon of the cavity (both at frequency $\omega_a$) to be converted into 1 photon of the buffer (at frequency $\omega_b$) and 1 photon of the microwave pump (at frequency $2\omega_a - \omega_b$). Since this buffer photon is then emitted to the environment, the cat qubit cavity is effectively subject to single-photon dissipation events, thus inducing exact parity swaps on the memory.

During a $Z(\theta)$ gate, we thus have the following situation. First, if no buffer photons are detected during the process, the system follows the dynamics of the no-jump deterministic equation, \ie~\eqref{eq:sme} with $d N_\eta = 0$. During the gate, modes $\a$ and $\b$ get entangled, but after the gate, $\b$ asymptotically relaxes back to vacuum under the no-detection backaction. The cat qubit then gets back to a pure state in which its phase information has been perfectly preserved (in addition to the bit value, which is exponentially protected).

Alternatively, one or several buffer photons can be detected during the process. This time, the system also follows the no-detection dynamics, up until the first detection event, at which point it is projected according to $\vrho \rightarrow \b \vrho \bdag$. This corresponds to a phase jump of approximately $\pi$, with the exact angle depending on jump time and gate parameters. This can be roughly understood by recalling the analysis of Section~\ref{sec:gateerrors}, where for an approximate model and in absence of photodetector, we observed that $\b(t)$ in the Heisenberg picture is proportional to $\sz$. After a detection-induced jump, we can thus perform classical feedback on the qubit to correct for this $\pi$ phase shift, hence improving gate fidelities.

\subsection{Measurement strategy}

The choice of a photodetector on $\b$ --- instead of e.g. homodyne or heterodyne detection --- is motivated as follows. If $\eta=1$, then no information is lost to the environment and the combined qubit-buffer state remains pure at all times, such that the final cat-qubit state is pure. However, this does not automatically imply that we would be able to perfectly restore the qubit state \textit{before} measurement; indeed, like in a standard perfect measurement, if the detections contain information about the qubit state, then the complementary qubit information is scrambled by back-action. According to the analysis of Section~\ref{sec:gateerrors}, the buffer real quadrature contains information about the qubit being in $\ket{0_L}$ or $\ket{1_L}$, and thus measuring this quadrature would necessarily induce qubit phase decoherence. Therefore, we choose to measure the energy (photon number) of the buffer, which erases this qubit logical information from detection results and hence should imply preservation of the qubit phase (and bit value) for $\eta=1$. In other words, by measuring the photon number, we prevent that the environment would induce detrimental backaction due to measuring the real quadrature of the buffer.

This picture is in fact exact. Both the Hamiltonian $\H_{AB} + \H_Z$ and the dissipation in $\mathcal{D}[\b]$ 
commute with a joint $x$-axis conjugation of both phase spaces. On the output channel $\b$ this conjugation involves a minus sign, which a quadrature measurement could in principle detect, but when measuring $\bdag \b$ this sign strictly disappears from the equations. Then the output signal contains zero information about the logical bit value of the cat qubit; hence, for $\eta=1$, the phase information of the cat qubit must be perfectly preserved. More details on this invariance of the master equation under joint phase conjugation can be found in Appendix~\ref{sec-apdx:xaxisconjugation}.

\subsection{Jump and no-jump trajectories}

Before discussing the full performance of the design with classical feedback, we examine jump and no-jump trajectories of the scheme separately.

Figure~\hyperref[fig:PD-rabi-osc]{2(a)} shows a numerical simulation of $Z$-axis Rabi oscillations both for a zero-photon-detected trajectory (solid blue) and for a single-photon-detected trajectory (dashed blue), as well as for the standard Zeno gate (black). To evaluate the qubit $\sx$  expectation value when the system is not exactly in codespace, we take the photon-number parity of mode $\a$. For the standard Zeno design, gate errors accumulate over time as shown by the decreasing amplitude of oscillations. For the photodetection design however, the qubit phase converges to an indeterminacy of order $10^{-3}$ (not visible) --- reflecting the steady state entanglement of $\a$ and $\b$ modes during gate operation (see Fig.~\hyperref[fig:bufferdynamics]{1(b)}) ---, and then keeps oscillating without loss. The single-jump trajectory clearly shows a full dephasing of angle $\simeq \pi$ after the detection event at a random time $t = t_J$, and further keeps oscillating without phase loss. Figure~\hyperref[fig:PD-rabi-osc]{2(b)} shows the probability that a detection event occurs during a $Z(\pi)$ gate, as a function of gate time and assuming an ideal photodetector. This jump probability is evaluated as $p_\text{J} = 1 - \exp \big(-\int \langle d N_\eta \rangle (t)\big)$. The low jump probability ensures that trajectories with more than 2 or 3 detection events will be extremely rare. 

\begin{figure}[!t]
         \centering
         \includegraphics[width = \columnwidth]{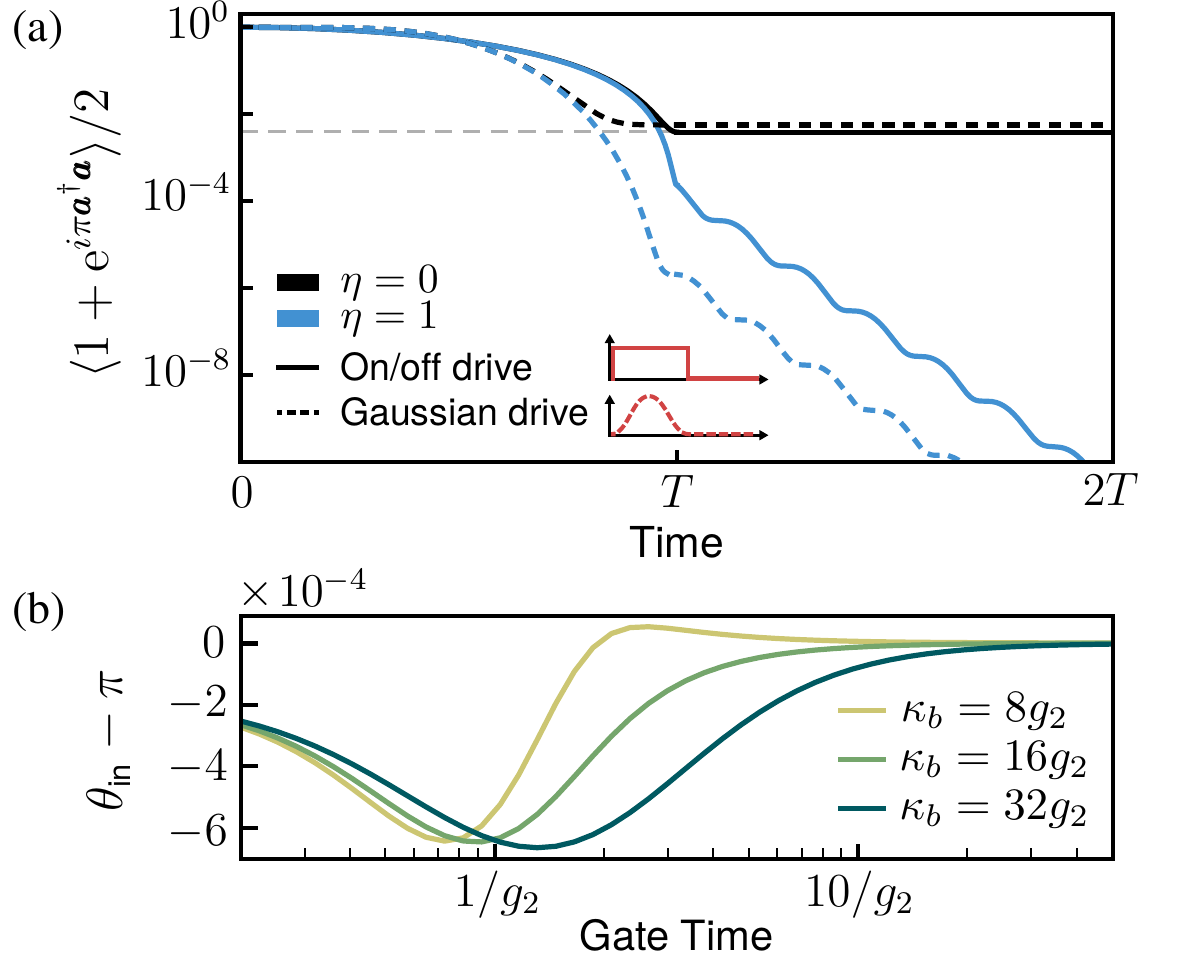}
    \label{fig:PD-conv}
    \vspace{-0.5cm}
    \caption{
        (a) Time-evolution of the parity during a $Z(\pi)$ gate with an ideal photodetector on the buffer mode output ($\eta = 1$) and without ($\eta = 0$). Only the no-detection trajectory is shown. The system is initialized in $\ket{+}_L \otimes \ket{0}$ and a single-photon drive~\eqref{eq:zdrive} is on for $0 < t < T$. For $t > T$, the buffer mode reconverges to vacuum as it disentangles from the cat qubit mode. The dashed gray horizontal line shows~\eqref{eq:pz0}. In this numerical simulation, $T = 4 / g_2$, $\kappa_b = 8g_2$ and $|\alpha|^2 = 8$. (b) Input angle $\theta_\text{in} = 4 \alpha \int \varepsilon_Z dt$, such that an exact $Z(\pi)$ gate is produced after in infinite-time reconvergence to the buffer mode vacuum.
    }
\end{figure}

The perfect preservation of cat qubit phase by gate operation holds after full disentanglement of the memory and buffer modes, which includes a reconvergence phase to the cat-qubit codespace after the gate drive has been turned off while still monitoring $\b$ with the photodetector. This is illustrated on Fig.~\hyperref[fig:PD-conv]{3(a)}, for two time-dependent shapes of single-photon drives. For $t \in [0,T]$ i.e.~while the gate drive is on, the build-up of a $\pi$-angle Rabi oscillation is observed, with better phase precision in presence of a photodetector ($\eta = 1$, blue) than without detector ($\eta = 0$, black). Also, as already observed in~\cite{xu2021engineering}, the gaussian-like time-dependent drive offers significantly better performance because the system mode is closer to its drive-less steady state at the end of the gate. For $t > T$, the single-photon drive is turned off and the reconvergence begins. In absence of a photodetector, the qubit phase remains perfectly constant during reconvergence since photon-number parity is conserved by the two-photon dissipation dynamics. This is shown on Fig.~\hyperref[fig:PD-conv]{3(a)} for the black curves. In contrast, when keeping the photodetector on the $\b$ mode for $t>T$, the SME does not preserve photon-number parity due to the non-linear backaction terms featured in~\eqref{eq:smeops-jump}, and therefore it is indeed possible to further improve the phase precision as the modes progressively disentangle. The oscillations observed during this reconvergence are due to the damped harmonic oscillator behavior of the mode, as shown in~\eqref{eq:bufferdyn}. 

For $\eta=1$, in principle there is no limit to phase precision after full disentanglement. To check this, we have adjusted the drive amplitude numerically to perform an ideal $Z(\pi)$ gate at $t\rightarrow \infty$, as shown in Fig.~\hyperref[fig:PD-conv]{3(a)}. Then, we have checked that the same drive amplitude indeed performs rotations of the same angle whichever the initial cat qubit state. Figure~\hyperref[fig:PD-conv]{3(b)} shows the typical small correction to be applied on the drive amplitude for trajectories without measurement detections.

\subsection{Design performance}

We now describe the performance of the full design with buffer mode photodetection and classical feedback. To quantify this, we simulate the following master equation,
\begin{equation}\label{eq:sme-feedback}
    \begin{split}
    d \vrho = &-i \left[\H_{AB} + \H_Z, \vrho\right] dt  \\
        &+ \kappa_b \mathcal{D}_\eta[\b] \vrho \, dt + \mathcal{J}[\vb{Z}(\pi)\b] \vrho \, dN_\eta
    \end{split}
\end{equation}
which is the same as~\eqref{eq:sme} but with a jump operator in $\vb{Z}(\pi) \b$ that indicates a Pauli correction on the memory mode for every buffer photon detected. Note that this stochastic master equation does not correspond to any physical model. We only introduce it to explain and quantify the idea behind the design. The actual feedback should be applied separately from the photodetection, for instance in software before any non-Clifford gate or with a subsequent $Z(k\pi)$ gate where $k$ is the number of detected photons. Alternatively, erasure errors can be included in the model, in which case qubits with at least one buffer photon detected during gates are discarded. Hence, with this plethora of possible feedback strategies, we limit our study to~\eqref{eq:sme-feedback}.

\begin{figure}[!t]
         \centering
         \includegraphics[width = \columnwidth]{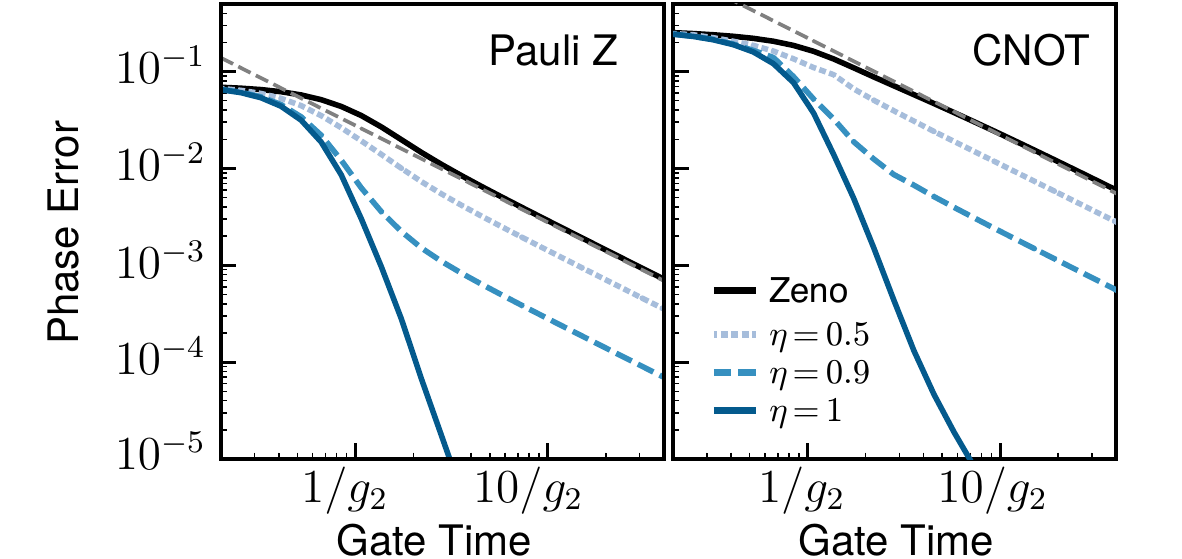}
\label{fig:PD-cnotNEW}
\vspace{-0.5cm}
\caption{
    Phase errors of $Z(\pi)$ and CNOT gates for the standard Zeno design ($\eta = 0$, black) and with a photodetector on the buffer mode output of the cat qubits (blue). The figure shows an average over every possible stochastic trajectory for the photodetection design of~\eqref{eq:sme-feedback}. Gate drives are Hamiltonians~\eqref{eq:zdrive} and~\eqref{eq:cnotff} with gaussian time-dependence. Dashed gray lines show~\eqref{eq:zgateerr} and~\eqref{eq:cnot-errs}. In these numerical simulations, $\kappa_b = 8g_2$ and $|\alpha|^2 = 8$.
}
\end{figure}

The $Z(\pi)$ gate performance achieved after a finite time with this design is shown on Fig.~\ref{fig:PD-cnotNEW}(a). The curves result from an average over several realizations for which the photodetector may have clicked at different times, and the drive is optimized for the zero-detection trajectory to achieve an ideal gate after full reconvergence to the codespace. For this feedback scheme, the gate fidelity is limited by finite detection efficiency as shown by the linear scaling of dashed blue lines that correspond to $\eta = 0.5$ and $\eta = 0.9$. For the ideal photodetector $\eta = 1$, gate fidelities scale with a high-order polynomial in the gate time. In principle, towards ultimate precision, one could perform an ideal feedback of angle $\phi(t_J) \approx \pi$ that depends on the exact jump time, and obtain an error-less gate. This is discussed further in Appendix~\ref{sec-apdx:feedback}.

Generalization of the design to multi-qubit CNOT and Toffoli gates is quite straightforward. Considering only the simpler scheme of~\cite{gautier2022combined} for which target mode stabilization is turned off during the gate process, photodetection should be performed on the buffer mode output of the control qubit(s) only. The numerical performance of this design for the CNOT gate is shown on Fig.~\ref{fig:PD-cnotNEW}(b). Here, feedback is also assumed to be perfectly applied following every buffer mode photodetection, according to~\eqref{eq:sme-feedback}. Similarly to the single-qubit gate, we find several orders of magnitude fidelity improvement in the ideal photodetector case, and otherwise a fidelity improvement limited by detection efficiency. In the next section, we discuss non-ideal photodetectors in more details.

\subsection{Non-ideal photodetector}

A realistic photodetector is never ideal and features a finite detection efficiency $\eta$. In this case, only part of the information lost to the environment is retrieved, and the resulting gate features dynamics in between the two regimes $\eta=0$ and $\eta=1$. A lower bound on the error is then given by 
\begin{equation}\label{eq:pzeta}
    p_Z \gtrsim (1-\eta)\; p_Z^{(0)} ,
\end{equation}
where $p_Z^{(0)}$ is the phase error of the regular Zeno gate, for instance as given by~\eqref{eq:pz0} for the $Z(\theta)$ gate. Indeed, a fraction $(1-\eta)$ of the state would behave as in the absence of a photodetector. We denote this as approximate because, when the photodetector does click, we can herald a successful detection trajectory if such heralding is compatible with the rest of the architecture. 

\begin{figure}[!t]
         \centering
         \includegraphics[width = \columnwidth]{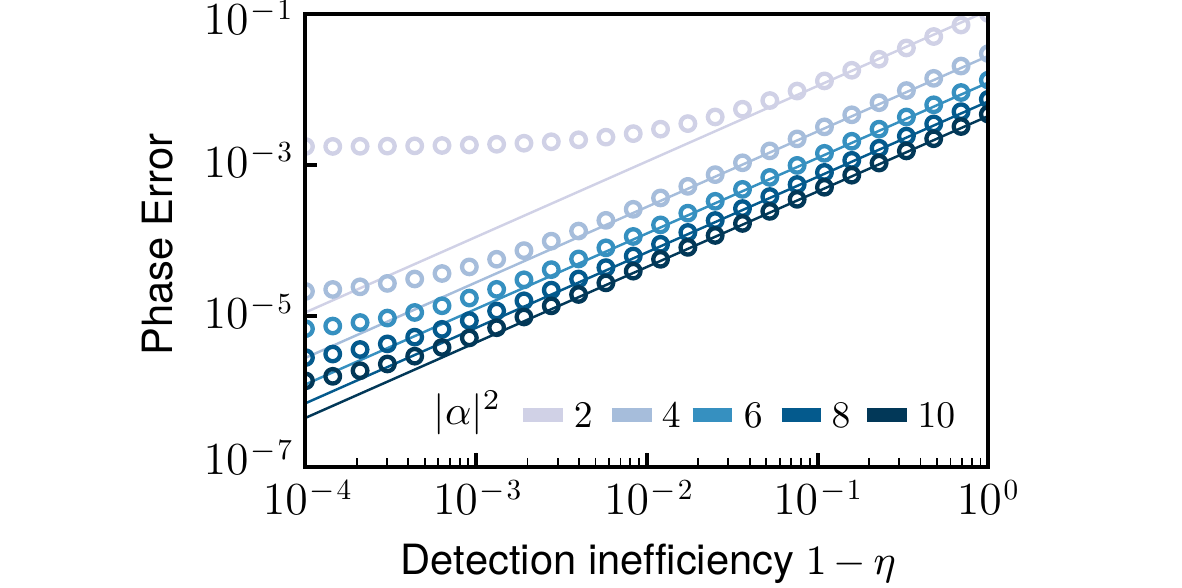}
    \label{fig:PD-eta}
    \vspace{-0.5cm}
    \caption{
        Scaling of phase errors of a Z($\pi$) gate against detection inefficiency $1-\eta$, for increasing values of $|\alpha|^2$. Gate errors are evaluated at $t = T = 4 / g_2$ without reconvergence ($T_c = 0$). Markers show numerical integration of the SME with classical feedback. Solid lines show~\eqref{eq:pzeta}, with a linear scaling with $1-\eta$. In this numerical simulation, $\kappa_b = 8 g_2$.
    }
 \end{figure}

Figure~\ref{fig:PD-eta} shows the numerical scaling of phase errors of a $Z(\pi)$ gate against the detection inefficiency $1-\eta$ in log-log scale. We indeed find a linear scaling with the amount of information lost to the environment according to~\eqref{eq:pzeta}, that eventually saturates for large enough detection efficiencies. This saturation results from the imperfect reconvergence to the codespace, as previously discussed.

In practice, a non-ideal photodetector does not only feature a non-unity detection efficiency, it may also have a finite dark count rate (the detector clicks without any photon measured) and a finite uncertainty on the detection time. The former would result in the erroneous application of a $\pi$-angle feedback at the dark count rate similar to thermal photons in the buffer mode (see Appendix~\ref{sec-apdx:thermalbuffer}). The latter has no impact on the design fidelity since a $Z(\pi)$ correction gate should be applied independently of the detection time. In any case, both of these effects are widely negligible compared to detection inefficiency, which even with state of the art photodetectors is typically in the $\eta = 0.1 \mbox{---} 0.5$ range. With the rapid improvement in photodetector efficiencies~\cite{albertinale2021detecting,dassonneville2020number,lescanne2020irreversible}, we may however expect this gate design to become viable in the coming years (see also Fig.~\hyperref[fig:errormodel]{13(a)}). The design can also inspire other feedback methods in which feedback is hardware-efficient or made autonomously, such as the one introduced in the following section.


\section{Cat-Buffer Autonomous Feedback}\label{sec:autonfeedback}
\subsection{Design principle}
This second gate design, inspired by the previous one, introduces a correlated dissipator to remove entropy from the system instead of the standard single-photon dissipation on the buffer mode. The corresponding master equation to be engineered on the cavity-buffer system then reads
\begin{equation}~\label{eq:cohfeedb-masteq}
    \frac{d \vrho}{dt} = -i \left[ \H_{AB} + \H_Z, \vrho \right] + \kappa_{ab} \mathcal{D}[\a\b] \vrho
\end{equation}
with the main addition of a two-mode dissipation operator that was recently realized in~\cite{gertler2022experimental} in the context of pair-cat codes stabilization~\cite{albert2019pair}. The main idea behind this peculiar dissipation arises from the photodetector scheme of the previous section. When the cat-qubit leaks out of its codespace under the gate drive, the buffer mode is populated through the $\H_{AB}$ interaction, then inducing an eventual phase-flip on the cavity mode as explained in Section~\ref{sec:gateerrors}. The correlated dissipation thus ensures that whenever the buffer mode loses a photon (hence inducing a $Z(\pi)$ error on the logical cat qubit), a direct photon loss on the cavity mode is also produced thus switching the cat qubit parity a second time and correcting for the error autonomously.

Mathematically, this can be understood using the shifted Fock basis~\cite{chamberland2020building} transformation as defined in~\eqref{eq:sfb-transf}, which yields
\begin{equation}\label{eq:cohfeedb-sfb}
    \begin{split}
        \kappa_{ab} \mathcal{D}[\a\b] &\rightarrow \kappa_{ab} \mathcal{D}[\sz (\ta + \alpha) \b] \\
        &= |\alpha|^2 \kappa_{ab} \mathcal{D}[\sz \b] + \mathcal{O}\left(|\tadag \ta|^{1/2}\right)
    \end{split}
\end{equation}
where the second line approximation holds since $|\tadag \ta| \ll 1$ in the limit of a small amount of leakage. Keeping only the leading order term in~\eqref{eq:cohfeedb-sfb}, and taking $\kappa_b \equiv |\alpha|^2 \kappa_{ab}$, it is possible to perform the same gate error derivation as in Sec.~\ref{sec:gateerrors}. This yields the exact same resulting set of equations on $\a$ and $\b$, but with a different dissipation operator in $\sz \b \propto \sz^2 = \vb*{I}$ where $\vb*{I}$ is the identity on the two-level cat qubit mode; \ie~the environment does not receive any information about the qubit state, as required. We emphasize once again that this derivation is only first-order and that non-linear effects have been neglected, for instance with terms in $\ta^2 \bdag + \hc$ that may eventually limit gate performances.

To engineer the correlated dissipation~\eqref{eq:cohfeedb-masteq}, an ancillary low-Q reservoir mode $\r$ can be introduced into the setup~\cite{gertler2022experimental}. By engineering four-wave mixing between those three modes and a classical pump, it is then possible to enable Hamiltonian interaction of the form $g_{ab} \a\b \rdag + g_{ab}^*\adag\bdag \r$. Together with a large reservoir damping of the form $\kappa_r \mathcal{D}[\r]$ and in the limit of $\kappa_r \gg g_{ab}$, it is possible to eliminate the fast dynamics of the reservoir mode. On the reduced system of the cat and buffer modes, the required correlated dissipator is then obtained, with typical amplitude $\kappa_{ab} = 4 g_{ab}^2 / \kappa_r$~\cite{azouit2017structure}.

In the limit of $\kappa_{ab} |\alpha|^2 \gg g_2$, it is further possible to eliminate the fast dynamics of the buffer mode to obtain an effective single-mode master equation on the cat qubit. Using the effective operator formalism of~\cite{reiter2012effective} yields the following dynamics on the cat qubit mode,
\begin{equation}\label{eq:cohfeedb-catdyn}
    \frac{d \vrho}{dt} = \frac{4g_2^2}{\kappa_{ab}} \mathcal{D}\left[\a (\adag \a)^{-1} (\a^2 - \alpha^2)\right] \vrho
\end{equation}
where $(\adag \a)^{-1}$ is the pseudo-inverse of the photon number operator and describes the effective difference in dynamics undergone during transient buffer excitation. This master equation indeed describes a parity-switching stabilization of the cat mode, but it differs from the master equation targeted in~\cite{xu2022autonomous}. More details on this model reduction can be found in Appendix~\ref{sec-apdx:adiabelim}.

\subsection{Parity-switching dynamics}

This dissipation with $\mathcal{D}[\a\b]$ is activated more generally when the cat-qubit leaks out of its codespace, triggering a reaction of its buffer. The effect will be beneficial whenever the leakage source is associated to parity-switching. For standard cat qubits, this is mainly the case for the user-induced Zeno dynamics as just discussed (for $Z(\theta)$ gates, on the control qubit of CNOT gates, or other similar gates). It can also occur by thermal excitation of the cat mode, of the form $\mathcal{D}[\adag]$. In contrast, leakage that preserves the photon-number parity would then induce phase errors by following the same process. This is for instance true of pure dephasing, of the form $\mathcal{D}[\adag \a]$. However, the amplitude of such effects is often negligible compared to other sources of phase errors such as gates and finite resonator lifetime.

The recent preprint of~\cite{xu2022autonomous} also explores how single-photon losses, of the form $\mathcal{D}[\a]$, can induce parity-switching leakage on a squeezed cat qubit. Indeed, while the annihilation operator leaves the coherent states constituting regular cats in place, it does induce leakage on squeezed coherent states~\cite{schlegel2022quantum}. As such, similarly to~\cite{xu2022autonomous}, the correlated dissipator introduced in this work performs an autonomous correction of single-photon annihilation on squeezed cats, to first order; a better correction would be obtained with dissipation in $\mathcal{D}[S(\a)\b]$ where $S(\a)$ is the squeezed annihilation operator, featuring the squeezed coherent state as an eigenstate. 

One of the main limitations of such approaches stems from thermal noise in the buffer mode, since any buffer excitation would decay by triggering a phase flip on the cat qubit. This effect is further discussed in Appendix~\ref{sec-apdx:thermalbuffer}.

\subsection{Design performance}
\begin{figure}[!t]
         \centering
         \includegraphics[width = \columnwidth]{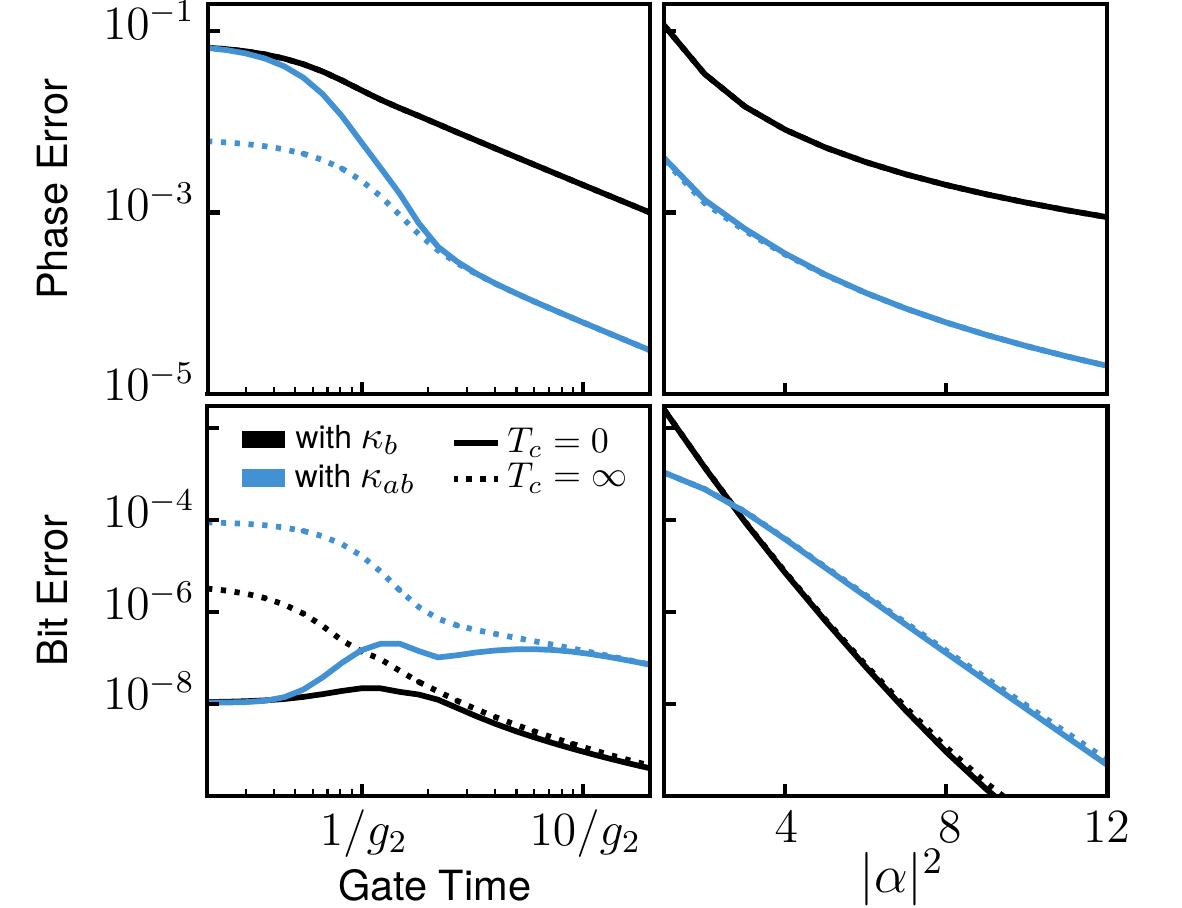}
\label{fig:cohfeedb-zgate}
\vspace{-0.5cm}
\caption{
Gate errors of a $Z(\pi)$ gate with the autonomous feedback design of~\eqref{eq:cohfeedb-masteq} at $\kappa_{ab} / |\alpha|^2 = 8 g_2$ (blue) and with the regular Zeno design at $\kappa_b = 8 g_2$ (black). Left: fixed cat size, $|\alpha|^2 = 8$. Right: fixed gate time, $T = 10 / g_2$. Gate errors are evaluated at $t=T$ or after full reconvergence to the steady state ($T_c=\infty$). While the cat qubit is entangled with its buffer and leaks out of codespace, the logical phase value is evaluated with photon-number parity on $\a$ and the logical bit value is evaluated with two-photon dissipation invariant~\cite{mirrahimi2014dynamically}.
}
\end{figure}

Let us now focus on the performance of this gate design, starting with the single-qubit $Z(\theta)$ gate. Figure~\ref{fig:cohfeedb-zgate} shows the errors induced by a $Z(\theta)$ gate for the regular Zeno gate (black) and for the correlated dissipator design of~\eqref{eq:cohfeedb-masteq} (blue) for both phase-flip and bit-flip errors. Following our previous analysis, the comparison between both designs is made at fixed $\kappa_b = |\alpha|^2 \kappa_{ab} = 8 g_2$ in order to keep the same damping rate. Left-side plots show the scaling with the gate time, while the right-side is plotted against cat size. For phase-flip errors, an improvement by a constant factor of about $\mu \approx 0.02$ is found, independent of both gate time and cat size, and limited by second-order effects as discussed previously. In the regime of short gate times, the gate drive is large in amplitude, and so the buffer is largely entangled with the cat mode at the end of the gate. For this reason, a reconvergence time is required to reach the constant fidelity gain of $\mu$. In other words, the buffer has not had enough time to lose its excitations to the environment, and so the correlated dissipator has not yet corrected for the coherent errors that occurred during the gate.

\begin{figure}[!t]
         \centering
         \includegraphics[width = \columnwidth]{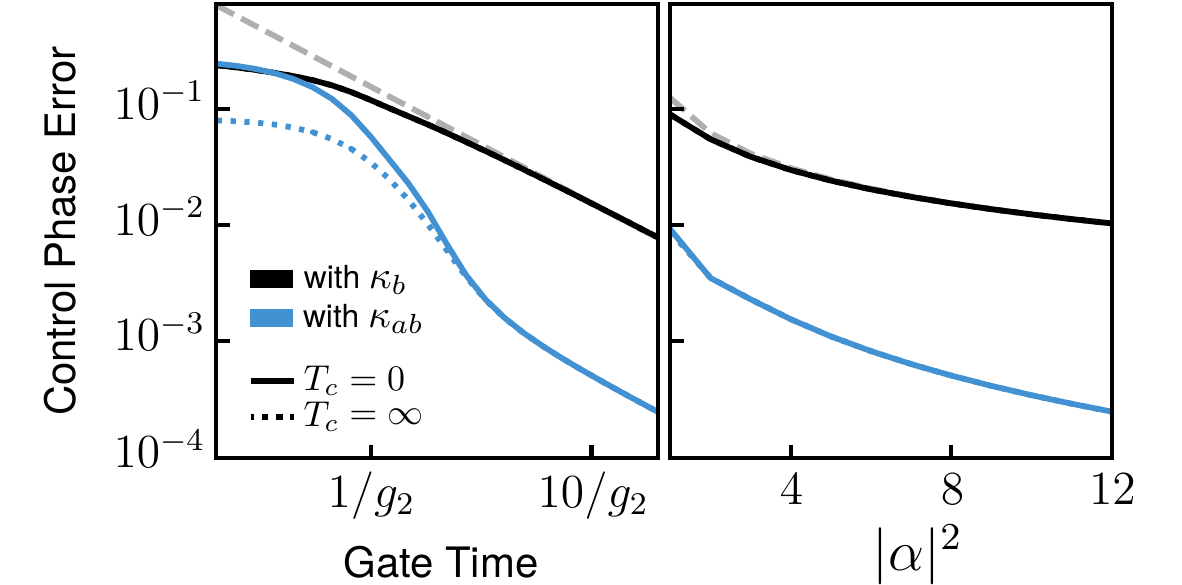}
\label{fig:cohfeedb-cnot}
\vspace{-0.5cm}
\caption{
Gate errors of a CNOT gate with the autonomous feedback design of~\eqref{eq:cohfeedb-masteq} at $\kappa_{ab} / |\alpha|^2 = 8 g_2$ (blue) and with the regular Zeno design at $\kappa_b = 8 g_2$ (black). Left: fixed cat size, $|\alpha|^2 = 8$. Right: fixed gate time, $T = 10 / g_2$. Gate errors are evaluated at $t=T$ or after full reconvergence to the steady state ($T_c=\infty$). Dashed gray lines show~\eqref{eq:cnot-errs}.
}
\end{figure}

The correlated dissipator of~\eqref{eq:cohfeedb-masteq} can raise concerns about bit-flip errors since it would, by itself, stabilize the vacuum state in both resonators. Hence, the bottom plots of Fig.~\ref{fig:cohfeedb-zgate} investigate this bit-flip error and shows that the exponential scaling in $|\alpha|^2$ is preserved, although slightly degraded from that of the regular Zeno gate. This is likely due to the additional leakage induced by the correlated dissipator. Indeed, whenever the buffer mode is populated, the cat mode is pushed towards its own vacuum state, at a rate proportional to the buffer population. When stopping operation before reconvergence (solid blue curve), the bit-flip rate looks artificially reduced for very short gate times, as the buffer had no time to dissipate yet.

\subsection{Multi-qubit gates}
For CNOT and Toffoli gates, the gate drive is parity-preserving on target qubit. As explained for the buffer photodetection gate design, in principle then detecting all buffer relaxations would induce a combination of phase corrections on the control qubit; to first order however, this comes down to only requiring the correlated dissipation on control qubits. Figure~\ref{fig:cohfeedb-cnot} investigates the performance of such a setup for two-qubit CNOT gates. The correlated dissipator of~\eqref{eq:cohfeedb-masteq} is activated on the control qubit, and the drive Hamiltonian of~\eqref{eq:cnotff} is switched on. Here, the performance is very similar to that of the single-qubit $Z(\theta)$ gate, with a phase fidelity improvement of $1/\mu \approx 50$ in the best case scenario. This is obtained at large gate times when the drive is small compared to the damping rate, $\varepsilon_{CX} \ll \kappa_{ab} |\alpha|^2$.


\section{Locally Flat Hamiltonian}\label{sec:flathamil}

\subsection{Intuition}

Quantum states feature a dispersion in position and momentum in their phase-space representation, with an equal variance in the case of coherent states. A single-photon drive, such as the gate drive of~\eqref{eq:zdrive}, thus acts differently along this dispersion and according to the specific position and momentum values of the state. This induces a phase-space displacement that can be frozen by a continuous measurement, such as the one of two-photon dissipation. This is the so-called Zeno effect.

From the wavefunction perspective, a coherent state is gaussian in position representation~\footnotemark[1] and, for a real coherent amplitude, reads $\psi(x) \propto \exp[-(x-\langle x \rangle)^2/2]$. Since the Zeno drive of~\eqref{eq:zdrive} reads $H_Z(x) = \varepsilon_Z x$ in this same representation, the time-evolution of the wavefunction under $H_Z$ can be trivially obtained as after some time $t$, the wavefunction reads $\psi(x,t) \propto \exp[-i x \varepsilon_Z t] \psi(x)$. This corresponds to a position-dependent phase shift, and therefore a displacement along the momentum axis $p$. However, two-photon dissipation prevents this displacement, and only the average phase shift $\exp[-i \langle x \rangle \varepsilon_Z t]$ remains, thus driving a cat qubit $Z(\theta)$ gate together with a logical phase blurring reflecting the variance of $\exp[-i x \varepsilon_Z t]$ over each coherent state.

From this viewpoint, one way to improve the precision of the $Z(\theta)$ gate is to use a drive Hamiltonian with little dispersion over each coherent state. Moreover, this property should be robust to all effects which the cat-qubit is meant to cover, i.e.~most prominently local displacements in phase space. Conversely, such a Hamiltonian would induce almost no displacement of the coherent states, nor deformation of any kind. Hence the phase gate could be implemented without relying on two-photon dissipation, and the evolution can be purely unitary which is another way to see that the gate would induce no phase losses.

\footnotetext[1]{Here we use the mathematical equivalence with a mechanical harmonic oscillator, to call position $x$ and momentum $p$ the real and imaginary quadratures of the electromagnetic mode.}

\subsection{Design principle}

\begin{figure}[!t]
         \centering
         \includegraphics[width = \columnwidth]{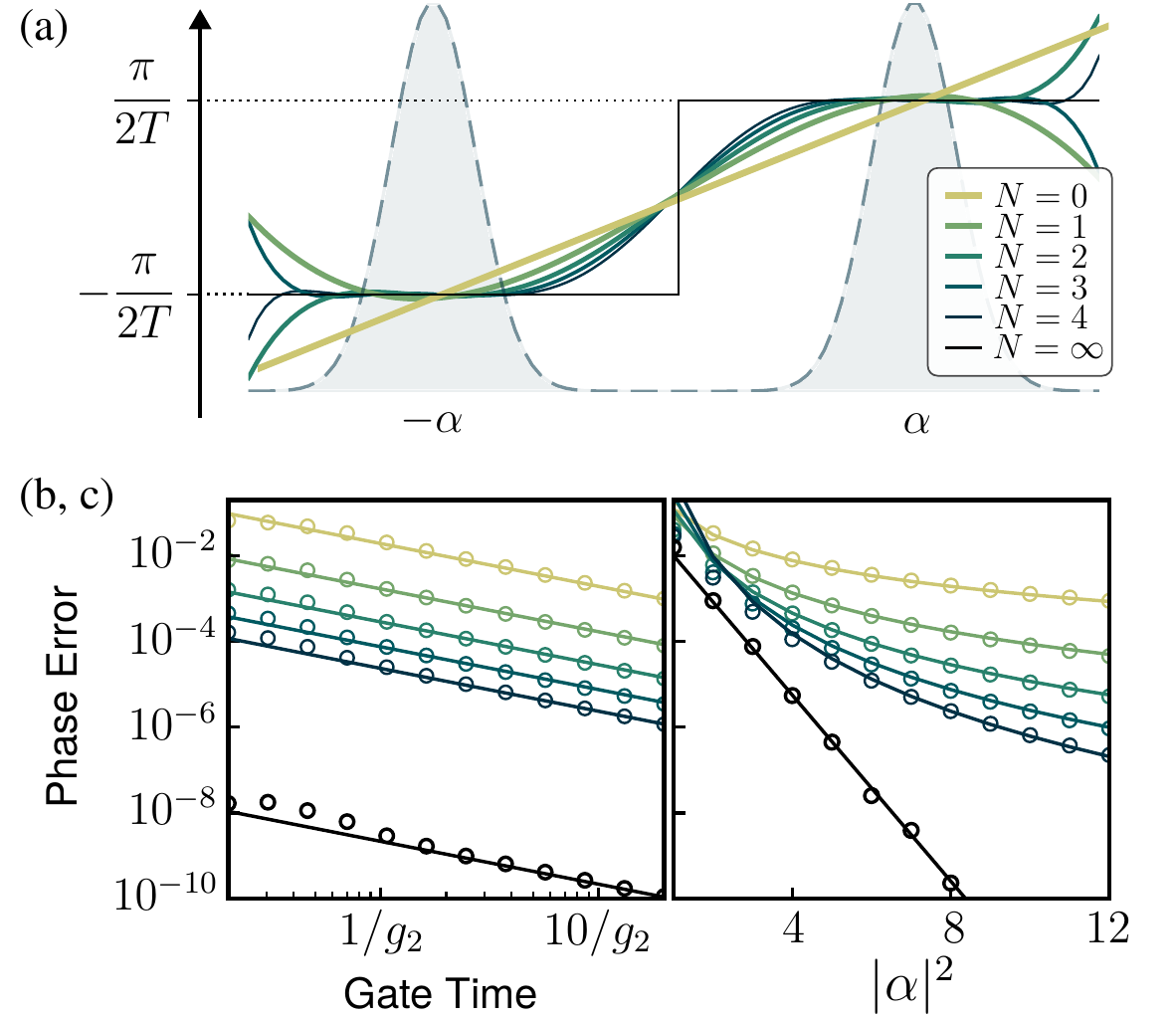}
\label{fig:polyn-zgate}
\vspace{-0.5cm}
\caption{
(a) Position distribution of locally flat drive Hamiltonians of~\eqref{eq:polyn-inf} and~\eqref{eq:polyn-N} that minimize the variance over the quantum fluctuations of cat states. (b, c) Gate phase errors for a $Z(\pi)$ gate with a locally flat drive Hamiltonian at (b) fixed cat size, $|\alpha|^2 = 8$, and (c) fixed gate time, $T = 10 / g_2$. In these simulations, $\kappa_b = 8 g_2$. Lines show numerical fits of the form $p_Z \propto T^{-1} |\alpha|^{-2(2 + N)}$ (color) and $p_Z \propto T^{-1} \exp(-2|\alpha|^2)$ (black). Markers show numerical data.
}
 \end{figure}

In this sense, an ideal~\footnotemark[2] drive Hamiltonian for single-qubit $Z(\theta)$ gates reads
\begin{equation}\label{eq:polyn-inf}
    \H_{Z,\infty} \equiv \varepsilon_Z \sign(\x)
\end{equation}
where $\sign$ denotes the sign function and $\x = \a + \adag$. This Hamiltonian yields a global phase difference for each half-plane of phase space, and thus engineers the required gate without any phase loss while being robust to any local error --- \eg~small displacements or distortions of cat states. It is represented in thin black lines on Fig.~\hyperref[fig:polyn-zgate]{8(a)}, together with the position representation of a superposition cat state in the background. Figures~\hyperref[fig:polyn-zgate]{8(b)} and~\hyperref[fig:polyn-zgate]{8(c)} show the phase errors induced by such a single-qubit $Z(\pi)$ gate with the Hamiltonian of~\eqref{eq:polyn-inf} in black lines, both for varying gate time (left) and varying cat size (right). We indeed find that this Hamiltonian provides phase errors that scale exponentially in the cat size, and that are drastically smaller than with the regular Zeno Hamiltonian of~\eqref{eq:zdrive}, represented in blue lines. 

\footnotetext[2]{This drive Hamiltonian is ideal up to exponentially small corrections, due to the orthonormalization of the computational basis --- two coherent states are never exactly orthogonal --- or in other words, to the gaussian tails of coherent states that cross the $x = 0$ line in their phase space distributions.}

The Hamiltonian of~\eqref{eq:polyn-inf} is highly non-linear in $\x$ and is therefore not accessible for state-of-the-art superconducting circuits, nor for other quantum computation platforms. It can however be approximated. Let us define the set of Hamiltonians
\begin{equation}\label{eq:polyn-N}
    \H_{Z, N} \equiv \varepsilon_Z \sum_{n=0}^N c_n \x^{2n+1}
\end{equation}
for $N \geq 0$, where $c_n$ are constants to be determined. These Hamiltonians have an odd distribution in position space. With the proper definition of $c_n$, they can approximate the sign Hamiltonian of~\eqref{eq:polyn-inf} locally around each cat qubit coherent state, even if the approximation cannot be global. The optimal $c_n$ constants should therefore minimize the gate errors induced by the combination of two-photon dissipation and drive during a given gate. Loosely speaking, these errors scale with the amount of non-flatness of the drive Hamiltonian of~\eqref{eq:polyn-N} --- a perfectly flat drive Hamiltonian would not displace the state and would thus result in an error-free gate. Therefore, we minimize the variance of the Hamiltonian across a single coherent state, given by
\begin{equation}\label{eq:polyn-var}
    V_N\left(\{c_n \}\right) \equiv \frac{1}{\sqrt{2\pi}}\int_{-\infty}^\infty H_{Z,N}(x)^2 \e^{-\frac{1}{2}\left(x-2\alpha\right)^2} d x ,
\end{equation}
under the constraint of a fixed angle of gate rotation,
\begin{equation}\label{eq:polyn-mean}
    \frac{1}{\sqrt{2\pi}}\int_{-\infty}^\infty H_{Z,N}(x) \e^{-\frac{1}{2}\left(x-2\alpha\right)^2} d x = \frac{\theta}{2 T}
\end{equation}
where $H_{Z,N}(x)$ is the position distribution of Hamiltonian~\eqref{eq:polyn-N}. This optimization problem is solved numerically using a Lagrange multiplier which is differentiated analytically and then minimized through matrix inversion. More details on this minimization can be found in Appendix~\ref{sec-apdx:polyn}.

Figure~\hyperref[fig:polyn-zgate]{8(a)} shows the first five of these Hamiltonians in colored lines, as determined by the previously-described minimization process. While the $N=0$ corresponds to the regular Zeno Hamiltonian of~\eqref{eq:zdrive} with a linear distribution, the $N > 0$ Hamiltonians show locally flat distributions around both coherent states, with increasing flatness as $N$ grows. 

\subsection{Results}
The performance of these Hamiltonians is then evaluated for the single-qubit $Z(\pi)$ gate in Figs.~\hyperref[fig:polyn-zgate]{8(b)} and \hyperref[fig:polyn-zgate]{8(c)}, with the gate phase error as a function of gate time and cat size respectively. The first plot shows a constant improvement in phase fidelity of the gate as $N$ increases, while the scaling with the gate time stays linear. The second plot however shows an improved scaling of phase errors with the cat size as $N$ grows. Although we lack an analytical derivation of phase errors for this set of drive Hamiltonians, a numerical fit of the form $p_Z \propto |\alpha|^{-2(2+N)}$ is performed as highlighted by the colored lines, and matches the numerical simulations particularly well in the large cat size limit. We attribute this scaling to the fact that, as both coherent states come further apart in phase space, the variance minimization process can achieve flatter distributions with each additional degree of freedom provided by the increasing $N$.

\subsection{Multi-qubit gates}
Generalization of this single-qubit gate design to multi-qubit gates is quite straightforward. Similarly as for the regular Zeno gate, the single-qubit drive of~\eqref{eq:polyn-N} should be multiplied by a phase-space rotation on the target qubit, which yields
\begin{equation}\label{eq:polyn-cnot-ff}
    \H_{CX, N} \equiv \varepsilon_{CX} \left(\sum_{n=0}^N c_n \x_C^{2n+1}  \right) \otimes \left( \adag_T \a_T - n_p \right)
\end{equation}
where $\x_C = \a_C + \adag_C$, the $\a_{C/T}$ are annihilation operators on control and target qubits respectively, $n_p$ is any even integer close to $|\alpha|^2$. Together with a static two-photon dissipation on the control qubit, and optionally with the correlated two-photon dissipation on the target qubit~\cite{guillaud2019repetition}, this process achieves a CNOT gate.

\begin{figure}[!t]
         \centering
         \includegraphics[width = \columnwidth]{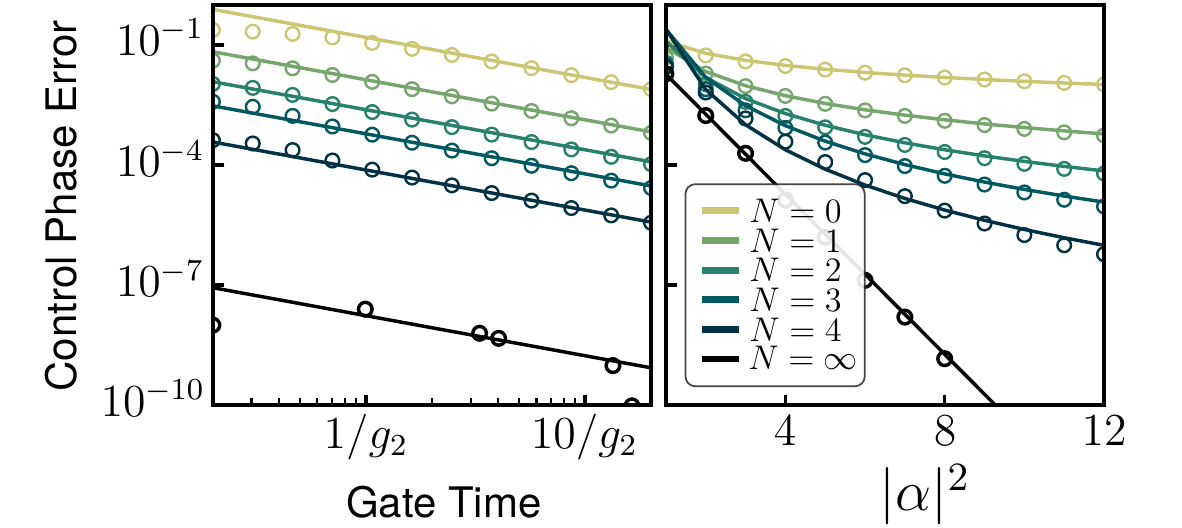}
\label{fig:polyn-cnot}
\vspace{-0.5cm}
\caption{
Gate-induced control phase errors for a CNOT gate with a locally flat drive Hamiltonian as in~\eqref{eq:polyn-cnot-ff}. Left: fixed cat size, $|\alpha|^2 = 8$. Right: fixed gate time $T = 10 / g_2$. Dissipative stabilization is acting on the control qubit only, with $\kappa_b = 8 g_2$. Lines show numerical fits of the form $p_{Z_C} \propto T^{-1} |\alpha|^{-2N}$ (color) and $p_{Z_C} \propto T^{-1} \exp(-2|\alpha|^2)$ (black). Markers show numerical data.
}
 \end{figure}

Figure~\ref{fig:polyn-cnot} investigates the phase errors induced by this two-qubit gate design on the control qubit, both against gate time (left) and cat size (right). Similar conclusions as for the single-qubit $Z(\theta)$ gate are reached. Gate errors scale linearly with time, and numerically we fit these errors according to $p_{Z_C} \propto |\alpha|^{-2(1+N)}$. The $|\alpha|^2$ difference in scaling compared to the single-qubit gate is due to the target qubit term in the drive Hamiltonian~\eqref{eq:polyn-cnot-ff}. Indeed, the operator $\adag_T \a_T - n_p$ induces a different (integer) number of $Z(\pi)$ gates on the control qubit for each Fock state of the target qubit, and the dispersion on this number increases with $\alpha$ (see~\cite{chamberland2020building} for details). 

\subsection{Engineering with an ATS}

Overall, the high-order Hamiltonians introduced in this section can greatly improve the performance of dissipative cat qubit gates, at the cost of additional difficulties to engineer the required Hamiltonians. Let us however note that high-order non-linearities are always present in superconducting circuits, even if they are often neglected due to their low amplitudes. Since the $c_n$ coefficients scale as $c_n \propto |\alpha|^{-2n}$, even moderate superconducting non-linearities could be enough to engineer these Hamiltonians, and so especially in the large cat size limit.

To illustrate this possibility, let us consider a realistic set of parameters for the current experimental proposal of dissipative cat qubits based on the Asymmetrically Threaded SQUID (ATS)~\cite{lescanne2020exponential,berdou2022one}. The ATS is a nonlinear circuit element made of a SQUID shunted by an inductance, which creates two flux loops that are then threaded at $0$ and $\pi$ flux bias respectively. The ideal Hamiltonian resulting from this setup reads
\begin{equation}
    \begin{split}
        \H &= \omega_a \adag \a + \omega_b \bdag \b \\
        &\quad - 2 E_J \varepsilon(t) \sin(\varphi_a (\a + \adag) + \varphi_b (\b + \bdag))
    \end{split}
\end{equation}
for an ATS coupled capacitively to both the cat qubit and buffer modes. Here, the $\varepsilon(t)$ term corresponds to a differential flux drive that can be frequency tuned to make specific terms in the sine Hamiltonian resonant. Also, $\varphi_{a/b}$ denote the energy participation of each mode into the ATS. Typically, it is desired to engineer $\varphi_a$ as large as possible to create strong two-to-one photon exchange rates, bearing in mind that a large $\varphi_a$ also induces increased high-order non-linear effects and increased single-photon losses by coupling to the buffer transmission line. 

For concreteness, let us assume that we want to implement a $Z(\pi)$ gate as in~\eqref{eq:polyn-N} with $N = 2$ (\ie~up to 5-th order Hamiltonian terms), with a gate time $T = 500\,\mathrm{ns}$ and for a cat of size $|\alpha|^2 = 8$. In this case, the global drive Hamiltonian amplitude to be engineered reads $\varepsilon_Z / 2 \pi= 1 / 8\alpha T \approx 88 \,\mathrm{KHz}$, and the polynomial coefficients of order 1, 3 and 5 read $c_0 \approx 0.66$, $c_1 \approx -0.055$ and $c_2 \approx 0.0021$ respectively. To make each of these coefficients match the ATS Hamiltonian terms, the following identities should be met,
\begin{subequations}\label{eq:coeffcond}
    \begin{align}
        \varepsilon_Z c_0 &= 2 E_J \varepsilon_0 \varphi_a + \mathcal{O}(\varphi_a^3) \\
        \varepsilon_Z c_1 &= 2 E_J \varepsilon_1 \varphi_a^3 / 3! + \mathcal{O}(\varphi_a^5) \\
        \varepsilon_Z c_2 &= 2 E_J \varepsilon_2 \varphi_a^5 / 5! + \mathcal{O}(\varphi_a^7)
    \end{align}
\end{subequations}
where $\varepsilon_k \ll 1$ are the flux drives amplitudes such that $\varepsilon(t) = \sum_k \varepsilon_k \cos((2k+1) \omega_a t)$. The right-hand side terms simply result from a Taylor expansion of the sine up to 5-th order in $\varphi_{a/b} \ll 1$. Further assuming realistic experimental parameters of $\varepsilon_k = 0.01$, $\varphi_a = 0.1$ and $E_J / 2 \pi = 90\,\mathrm{GHz}$~\cite{lescanne2020exponential}, we find that the right-hand side terms of~\eqref{eq:coeffcond} are at least 10 times larger than required by the left-hand side side terms. In other words, the achievable experimental parameters are 10 times larger than the actual drive amplitudes to be engineered. This gives some leeway to either implement even higher-order gate designs, or to release experimental constraints. In addition, for the same parameters, a factor of $|\alpha|^4 = 64$ improvement in gate fidelities can be expected, making the design particularly attractive.

As a final remark, one may also seek to engineer such nonlinear terms e.g.~with time-dependent schemes and rotating wave approximation, or by involving auxiliary systems. Although, such attempts should keep in mind the utmost importance of preserving the noise bias throughout gate operation.


\section{Discrete Jump}\label{sec:discrete}

For this final design, we shift away from the Zeno effect that has for now been the common basis for all physical cat gate implementations with a $Z$ component. This section introduces discrete gates that rely on a dissipative coupling to an ancillary mode. We begin with single-qubit $Z(\theta)$ gates, and generalize to CZ and CNOT gates afterwards.

\subsection{Design principle}

Consider the following master equation with the buffer mode adiabatically eliminated,
\begin{equation}\label{eq:discrete-zgate}
    \frac{d \vrho}{dt} = \kappa_2 \mathcal{D}[\a^2 - \alpha^2] \vrho + \kappa_Z \mathcal{D}[\a_\theta \sp] \vrho
\end{equation}
where $\sp$ denotes the creation operator of an ancillary qubit mode, and
\begin{equation}
    \a_\theta = \cos(\theta / 2) \alpha + i\sin(\theta / 2) \a
\end{equation}
is a modified annihilation operator such that $\a_\pi = i\a$, and $\a_\theta \ket{\pm \alpha} = \exp(\pm i \theta / 2) \ket{\pm \alpha}$. From this identity, it is immediate to verify that projecting any cat qubit state with $\a_\theta$ achieves a $Z(\theta)$ rotation of this state, up to exponentially small corrections in $|\alpha|^2$ to account for state normalization. Furthermore, assuming that this ancillary mode is initialized in its ground state $\ket{g}$ and does not suffer from errors, the correlated dissipator in $\a_\theta\sp$ engineers the loss of exactly one $\a_\theta$ photon from the cat mode, after which nothing more can happen with this dissipation. This can thus be seen as a kind of photon blockade technique, which can further be made error-resilient by, for instance, sending the additional photon to the meta-stable level of a three-level $\Lambda$-system or into an infinite transmission line.

For the particular value $\theta = \pi$,~\eqref{eq:discrete-zgate} induces the loss of exactly one $\a$ photon. Since the logical $\ket{\pm_L}$ cat states feature even and odd photon-number parities respectively and the logical bit value is exponentially protected under such losses thanks to the two-photon dissipation, this peculiar dissipator results in a Pauli $Z(\pi)$ gate on the cat qubit mode after infinite time evolution.

To estimate gate fidelities at finite times analytically, let us assume that the complete system is initialized in $\vrho(0) = \vrho_{A,0} \otimes \ketbra{g}{g}$ with $\vrho_{A,0}$ the initial density matrix on the cat mode. The full system density matrix at time $t$ can then be separated along the diagonal matrix elements of the ancillary mode according to $\vrho = \vrho_g \otimes \ketbra{g}{g} + \vrho_e \otimes \ketbra{e}{e}$ (the correlated dissipator of~\eqref{eq:discrete-zgate} does not induce diagonal to off-diagonal transitions). Inserting this expression in~\eqref{eq:discrete-zgate} thus yields
\begin{subequations} 
    \begin{align}
        \frac{d \vrho_g}{dt} &= \kappa_2 \mathcal{D}[\a^2 - \alpha^2] \vrho_g - \frac{1}{2}\kappa_Z \left\{ \adag_\theta \a_\theta, \vrho_g \right\} \label{eq:discrete-rhog}\\
        \frac{d \vrho_e}{dt} &= \kappa_2 \mathcal{D}[\a^2 - \alpha^2] \vrho_e + \kappa_Z \a_\theta \vrho_g \adag_\theta \label{eq:discrete-rhoe}
    \end{align}
\end{subequations}
where $\vrho_g(0) = \vrho_{A,0}$ and $\vrho_e(0) = 0$. Let us first discuss these coupled equations without the $\{\adag_\theta \a_\theta, \cdot\,\}$ term of~\eqref{eq:discrete-rhog}, which will be shown to be the main limitation of the design. As can be seen from the right-hand side term in~\eqref{eq:discrete-rhoe}, a transfer of population is made from $\ket{g}$ to $\ket{e}$ at rate $\kappa_Z |\alpha|^2$. Upon this population transfer, a single $\a_\theta$ projection of the cat mode is performed thanks to the $\a_\theta \vrho_g \adag_\theta$ term, thus achieving the required gate. The minimal gate-induced phase errors that can be achieved at finite time thus come down to the fraction of states which have not undergone the jump from $\ket{g}$ to $\ket{e}$, and read
\begin{equation}\label{eq:discrete-gateerr}
    p_Z = \exp(-|\alpha|^2 \kappa_Z t)
\end{equation}
where $t$ is the time of evolution under the correlated dissipator. The phase error would thus decrease exponentially with $|\alpha|^2$ and with the effective gate time $\kappa_Z t$.

Let us now consider the effect of the right-hand side term of~\eqref{eq:discrete-rhog}.

For the particular case $\theta = \pi$, since $\a_\pi = i \a$ and thus $\adag_\pi \a_\pi = \adag \a$, this term remains a parity-preserving operator. Therefore, the evolution with~\eqref{eq:discrete-rhog} and~\eqref{eq:discrete-rhoe} perfectly preserves the phase of the logical qubit, up to inducing the desired phase-flip gate from $\rho_g$ to $\rho_e$. Besides this, the last term of~\eqref{eq:discrete-rhog} acts much like a dispersive dissipation $\mathcal{D}[\a^\dagger \a]$, inducing noisy rotation of the cat-qubit out of its codespace. As long as $\alpha \kappa_Z \ll 4 |\alpha|^2 \kappa_2$, this effect is countered by the two-photon dissipation. Since the mean energies of even and odd cats are exponentially close, the induced logical bit-flip is exponentially small in $|\alpha|^2$. Up to an upper bound on $\kappa_Z/\alpha$, the exponentially scaling phase gate thus indeed holds.

For gate angles $\theta\neq \pi$, to consider the effect of the last term in~\eqref{eq:discrete-rhog} we write out
\begin{equation}\label{eq:discrete-parityswitch}
    \begin{split}
        \adag_\theta \a_\theta &= \cos^2\left(\theta / 2\right) |\alpha|^2 + \sin^2\left(\theta / 2\right) \adag \a \\
        & \quad - i \sin(\theta) (\adag - \a) / 2.
    \end{split}
\end{equation}
The second line of this operator would induce parity-switching leakage from the cat codespace, and therefore parity errors at a rate proportional to $\sin^2(\theta)$ once brought back to the codespace by two-photon dissipation. The gate fidelities obtained from this scheme are still competitive with other gate designs, and particularly in the limit of $\kappa_Z \ll \kappa_2$; see simulation results below. This can be understood as the last term in~\eqref{eq:discrete-rhog} has an effect roughly similar to a Hamiltonian in $i \sin(\theta) (\adag - \a)$. Hence, much like the analysis of Section~\ref{sec:gateerrors}, it induces phase errors at a rate proportional to $\kappa_Z^2/\kappa_2$, while the phase gate happens at rate $\kappa_Z$.

\subsection{Qutrit design}

But in fact, we can do better and retrieve for any $\theta$ the same performance as the $\theta=\pi$ case with a slightly more involved gate design. Consider the following master equation,
\begin{equation}\label{eq:discrete-zgate-qutrit}
    \begin{split}
    \frac{d \vrho}{dt} &= \kappa_2 \mathcal{D}[\a^2 - \alpha^2] \vrho + \kappa_Z \mathcal{D}[\a_\theta \ketbra{e}{g}] \\
    &\quad + \kappa_Z \mathcal{D}[\a_{\theta+\pi} \ketbra{f}{g}] + \kappa_Z' \mathcal{D}[\a_\pi \ketbra{e}{f}]
    \end{split}
\end{equation}
where $\ket{g}$, $\ket{e}$ and $\ket{f}$ denote the three lowest energy levels of a qutrit, for instance that of a transmon. This master equation now involves three population transfers. The first is the same as in~\eqref{eq:discrete-zgate} and performs a $Z(\theta)$ gate with a $\ket{g}$ to $\ket{e}$ transfer. The other two terms also describe a $Z(\theta)$ gate, but made in two steps by first transferring to $\ket{f}$ and then to $\ket{e}$. The main goal of adding these two terms is to cancel out parity-switching dynamics in the $\ketbra{g}{g}$ subspace. Indeed, we now have that
\begin{equation}
    \adag_\theta \a_\theta + \adag_{\theta + \pi} \a_{\theta + \pi} \propto \cos^2(\theta / 2) |\alpha|^2 + \sin^2(\theta / 2) \adag \a.
\end{equation}
In the $\ketbra{f}{f}$ subspace, the dynamics is also parity-preserving since a $Z(\pi)$ gate is performed. In other words, two paths have been constructed which both feed into the same final state with the required $Z$ rotation angle of $\theta$, and whose interference cancels the parity-switching term in~\eqref{eq:discrete-parityswitch}. This scheme would mainly be limited by qutrit characteristics, to be specified from experimental implementation, and by the requirement that both paths should feature the same rate of dissipation. In the following, we focus on the qubit-enabled $Z(\pi)$ gate, and also compare qubit- and qutrit-enabled $Z(\theta)$ gate designs.

\subsection{Results}

\begin{figure}[!t]
         \centering
         \includegraphics[width = \columnwidth]{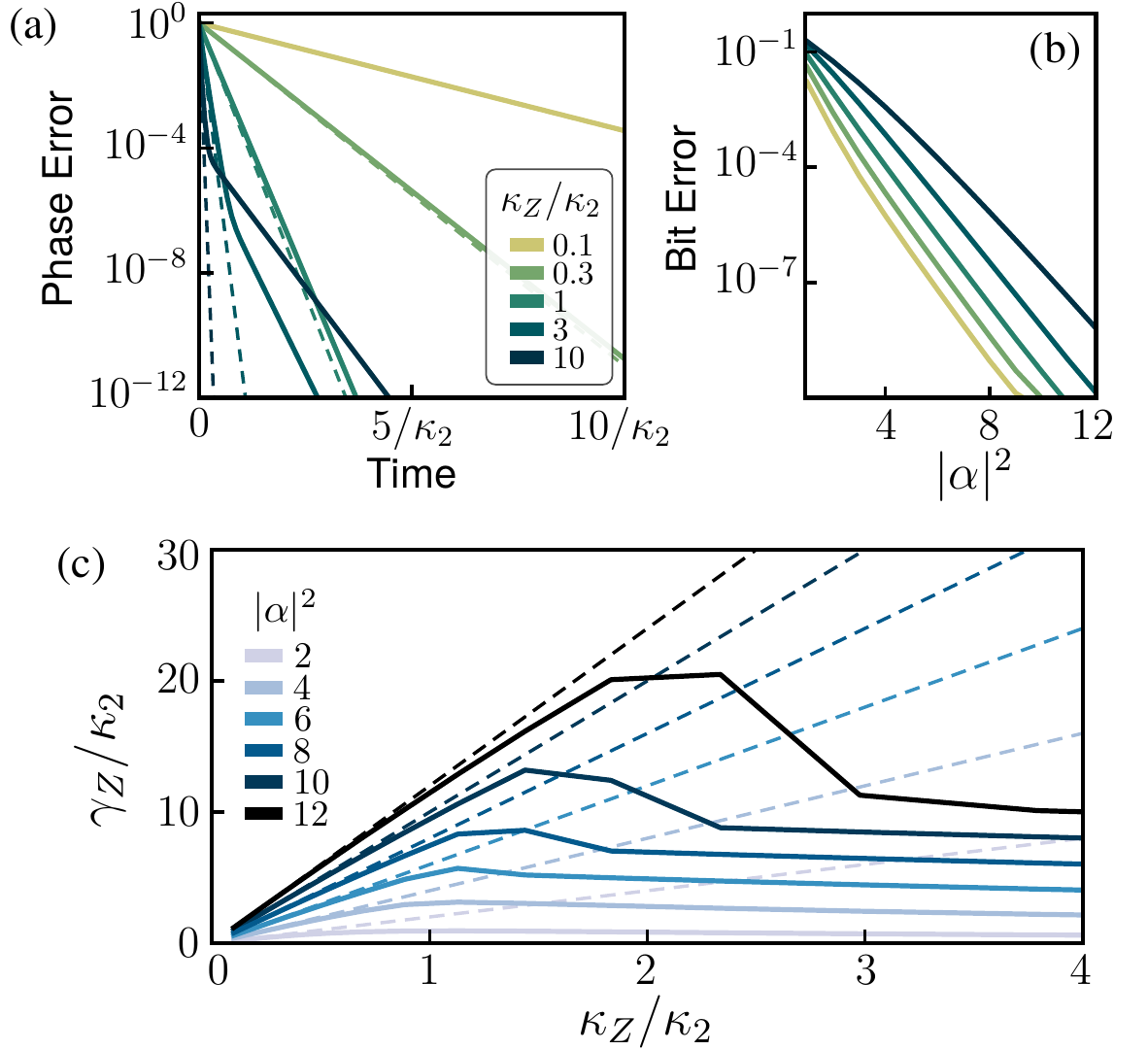}
\label{fig:discrete-zgate-fit}
\vspace{-0.5cm}
\caption{
(a,b) Gate-induced phase and bit errors of a discrete $Z(\pi)$ gate as in~\eqref{eq:discrete-zgate} for multiple values of $\kappa_Z / \kappa_2$. Left: fixed cat size, $|\alpha|^2 = 8$. Right: fixed gate time, $T = 10 / \kappa_2$. In the left plot, dashed lines show $p_Z = \exp(-|\alpha|^2 \kappa_Z t)$. (c) Exponential rate $\gamma_Z$ obtained from a linear fit of $\gamma(t) = -\ln(p_Z(t)) \propto \gamma_Z t$. Dashed lines show $\gamma_Z = |\alpha|^2 \kappa_Z$.
}
 \end{figure}

First, we investigate $Z(\pi)$ gates. Figure~\ref{fig:discrete-zgate-fit} shows the phase errors induced by this discrete $Z(\pi)$ gate without any additional error processes on the cat or ancillary modes. On plot (a), phase errors are shown in semi-log scale as a function of time for varying $\kappa_Z / \kappa_2$, and fixed $|\alpha|^2 = 4$ or $8$ in the master equation~\eqref{eq:discrete-zgate}. For each value of $\kappa_Z / \kappa_2$, both numerical simulations (solid) and the minimal gate error formula of~\eqref{eq:discrete-gateerr} (dashed) are shown. We retrieve that, in the limit of $\kappa_Z \ll \kappa_2$, the expected formula fits numerical simulations perfectly and exponentially small gate errors are achieved. As $\kappa_Z$ is increased, we find a deviation from the optimal gate errors due to the competition with two-photon dissipation, as the cat significantly leaves the codespace of intended size $|\alpha|^2$ during the gate, getting drawn closer to the vacuum. An exponential scaling with time is maintained, although with a smaller exponential rate, as discussed in the previous subsection.

In Fig.~\hyperref[fig:discrete-zgate-fit]{10(c)}, we extract this exponential scaling rate from a linear fit of $\gamma(t) = - \ln(p_Z(t))$, which yields $\gamma_Z$ such that $p_Z(t) \sim \exp(-\gamma_Z t)$ for sufficiently large time values. This exponential rate is then plotted as a function of $\kappa_Z / \kappa_2$ for varying values of $|\alpha|^2$, and also compared to the ideal rate of~\eqref{eq:discrete-gateerr} given by $\gamma_Z = |\alpha|^2 \kappa_Z$. Again, a transition from the optimal gate error regime to a sub-optimal regime is found as $\kappa_Z / \kappa_2$ is increased. In addition, the point of transition scales as $\alpha$, which would confirm the limit of validity of the optimal regime given by $\alpha \kappa_Z \ll 4 |\alpha|^2 \kappa_2$, or equivalently, $\kappa_Z / \kappa_2 \ll 4 \alpha$.

\begin{figure}[!t]
         \centering
         \includegraphics[width = \columnwidth]{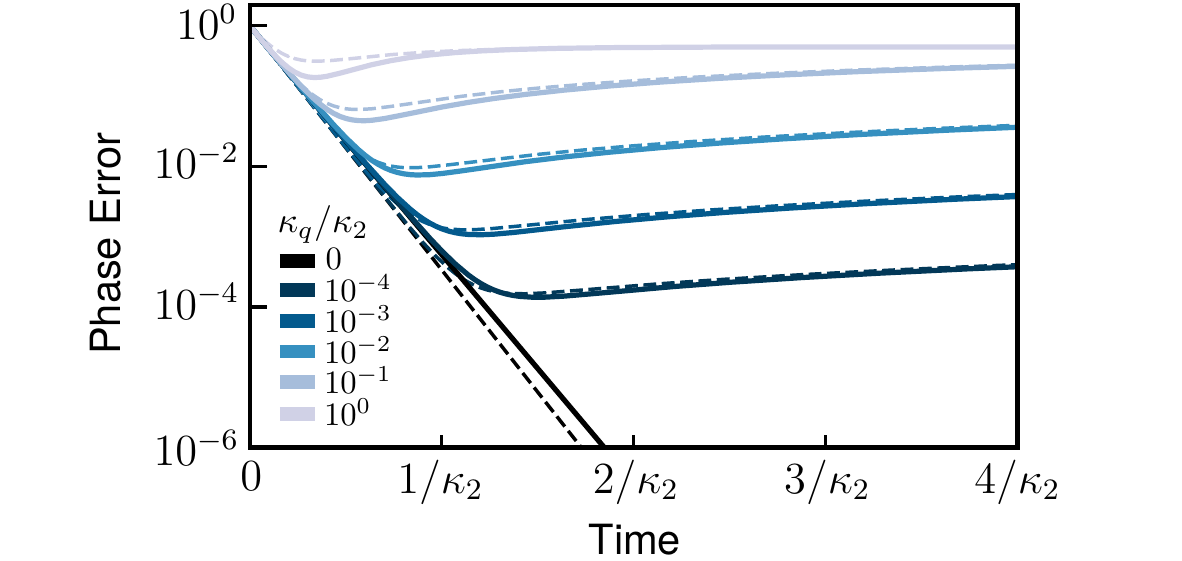}
\label{fig:discrete-qubiterr}
\vspace{-0.5cm}
\caption{
Gate-induced phase errors of a discrete $Z(\pi)$ gate as in~\eqref{eq:discrete-zgate} with additional single-photon losses of the form $\kappa_q \mathcal{D}[\sm]$ on the ancillary qubit. Dashed lines show $p_Z = \kappa_q t + \exp(-|\alpha|^2 \kappa_Z t)$. The cat size is fixed at $|\alpha|^2 = 8$ and $\kappa_Z / \kappa_2 = 1$.
}
 \end{figure}

The effect of $\kappa_Z / \kappa_2$ on \textit{bit-flip} errors is investigated on Fig.~\hyperref[fig:discrete-qubiterr]{10(b)}. In this particular simulation, a relatively large gate time is fixed at $T = 10 / \kappa_2$ to ensure the relevance of the study. We find that bit errors are indeed suppressed exponentially according to $p_X \propto \exp(-2|\alpha|^2)$ for all values of $\kappa_Z$ investigated, thus preserving the error bias of cat qubits. However, as the effective cat size is reduced by the $\kappa_Z$ term of~\eqref{eq:discrete-rhog} during the gate time, the prefactor of this exponential increases linearly with $\kappa_Z / \kappa_2$.

The main side-process limiting the fidelity of discrete gates would be ancillary mode lifetime. Since the gate is based on a transition from the ground to the excited state of the ancilla that will serve as a photon blocker, any unwanted transition between these two ancillary states will perturb the gate process. Figure~\hyperref[fig:discrete-qubiterr]{11} investigates finite qubit lifetimes for the two-level ancillary mode. Here, a discrete $Z(\pi)$ gate is simulated numerically in the presence of single-photon losses on the ancillary qubit of the form $\kappa_q \mathcal{D}[\sm]$, with varying values of $\kappa_q / \kappa_2$. An analytical fit is further shown for each numerical simulation, which reads $p_Z(t) = \kappa_q t + \exp(-|\alpha|^2 \kappa_Z t)$ and matches each line very well. Indeed, ancillary mode losses induce unwanted $\ket{e} \rightarrow \ket{g}$ transitions at a constant rate which are followed exponentially quickly by a $\ket{g} \rightarrow \ket{e}$ transition of the discrete gate correlated dissipator, inducing a second phase-flip and hence cancelling the $Z(\pi)$ gate. Thermal excitations of the ancillary mode would have a similar effect by activating the $\ket{g} \rightarrow \ket{e}$ transition without a cat mode parity switch, but in a cold environment such excitations have a much smaller rate than qubit decay. 

While this linear increase of phase errors may appear limiting for the usefulness of the gate design, we remind that finite lifetime of the cat oscillator also induce phase errors that scale linearly in time, according to $p_Z = |\alpha|^2 \kappa_1 t$ where $\kappa_1$ is the rate of single-photon losses. Those losses are inevitable without changing the overall cat qubit encoding. In contrast, for the ancilla, while a transmon-like qubit would yield a simple gate implementation, the particular gate design is compatible with more specific quantum systems. There is no requirement for the protection of the $\ket{g} + \ket{e} \leftrightarrow \ket{g} - \ket{e}$ transition, since only the diagonal elements are used by the gate design. As such, any ancilla system that can robustly implement a single transition from $\ket{g}$ to $\ket{e}$ would suffice for better gate protection. One can think of a three-level $\Lambda$ system with two meta-stable ground states~\cite{kumar2016stimulated,vepsalainen2019superadiabatic}, or a system where the ancilla state would escape away (but never back to $\ket{g}$) after reaching $\ket{e}$.

\begin{figure}[!t]
         \centering
         \includegraphics[width = \columnwidth]{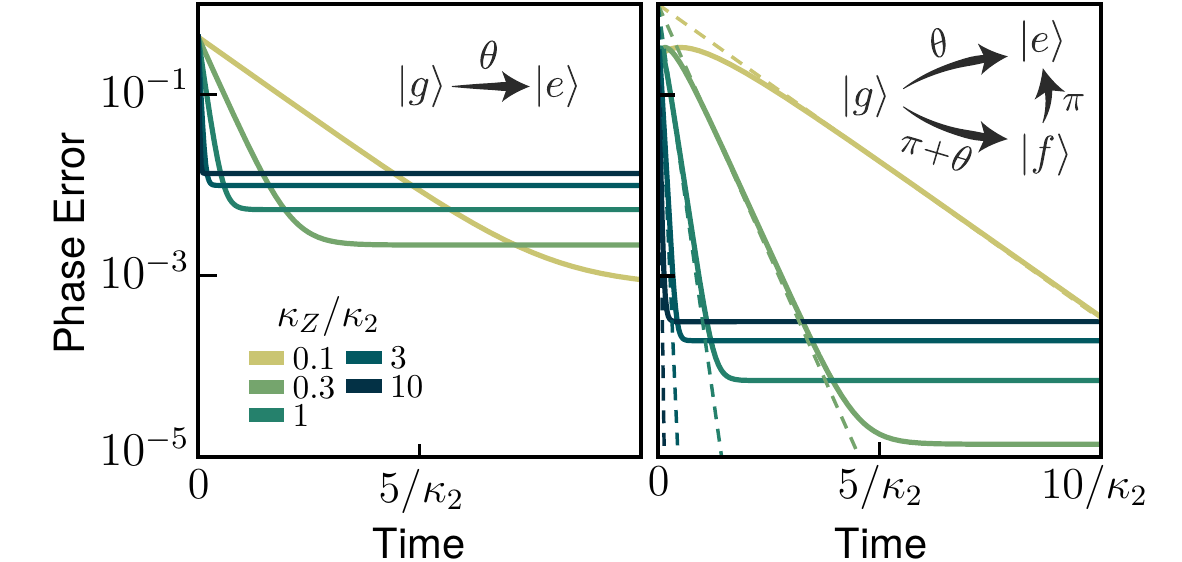}
\label{fig:discrete-angle}
\vspace{-0.5cm}
\caption{
Gate-induced phase errors of a discrete $Z(\pi/3)$, with the designs of~\eqref{eq:discrete-zgate} using an ancillary two-level system (left) and of~\eqref{eq:discrete-zgate-qutrit} using an ancillary three-level system (right). Dashed lines show $p_Z = \exp(-|\alpha|^2 \kappa_Z t)$. Insets show the gate transitions between ancillary states for either design.
}
 \end{figure}

Figure~\ref{fig:discrete-angle} investigates other gate angles, $\theta \neq \pi$, both with the ancillary two-level system design of~\eqref{eq:discrete-zgate} (left), and with the three-level system design of~\eqref{eq:discrete-zgate-qutrit} that cancels out parity-switching in the $\ketbra{g}{g}$ subspace through interference (right). In the first scheme, the phase errors first scale exponentially until they hit a plateau corresponding to the parity-switching leakage of~\eqref{eq:discrete-parityswitch}. This plateau scales according to $\sin^2(\theta)$. With the qutrit scheme, the exponential scaling of phase errors is also observed but for longer time scales, thus confirming that parity-switching dynamics is canceled out. A plateau is still hit after some time $t$ but at lower phase errors than for the previous two-level system scheme. This new limitation is explained by the fact that, if the state is initially not coherent or with a different coherent state amplitude than $\alpha$, $\a_\theta$ does not map exactly onto a $Z(\theta)$ gate. Since the $\ketbra{g}{g}$ subspace dynamics induces such incoherent fluctuations, gate fidelities are limited by this effect. For sufficiently small values of $\kappa_Z / \kappa_2$, the state however stays inside the codespace at all times and the exponential scaling of gate errors is maintained for longer time scales.

\subsection{Multi-qubit gates}
Regarding multi-qubit gates, the single-qubit discrete method of~\eqref{eq:discrete-zgate} can be generalized but not necessarily in a trivial manner. For CZ gates, consider the two-qubit operator,
\begin{equation}
    \L_{CZ} = - \a_1 (\a_2 - \alpha) + \alpha (\a_2 + \alpha)
\end{equation}
where $\a_{1/2}$ are the annihilation operators on each mode involved. This operator is such that $\L_{CZ} \ket{\pm \alpha}_1 \ket{\alpha}_2 = 4\alpha \ket{\pm \alpha}_1 \ket{\alpha}_2$ and $\L_{CZ} \ket{\pm \alpha}_1 \ket{-\alpha}_2 = \pm 4\alpha \ket{\pm \alpha}_1 \ket{\alpha}_2$ which thus meets the $CZ$ gate requirements. Note in particular the inter-exchangeability of $\a_1$ and $\a_2$ in this operator, which shows the $CZ$ gate symmetry. By engineering a dissipation of the form $\mathcal{D}[\L_{CZ} \sp]$, an exponentially scaling $CZ$ gate can be achieved. It would however suffer from the same issues as qubit-enabled $Z(\theta)$ gates due to the $\ketbra{g}{g}$ parity-switching dynamics, since $\L^\dagger_{CZ} \L_{CZ}$ contains parity-switching terms both in $\a_1$ or $\a_2$. Similarly as for single-qubit $Z(\theta)$ rotations, it is possible to cancel out these terms through interference by instead using four operators similar to $\L_{CZ}$ and a higher-dimensional ancillary system.

For CNOT gates, consider the two-qubit operator,
\begin{equation}\label{eq:discrete-cnot}
    \L_{CX} = \alpha \vb*{I} \otimes \vb*{P}_{+} + \a \otimes \vb*{P}_{-}
\end{equation}
where the first and second modes in this outer product are control and target qubits respectively, where $\vb*{I}$ is the oscillator identity, and $\vb*{P}_\pm = (\e^{i\pi \adag \a} \pm 1) / 2$ are projectors on the even and odd parity subspaces of the target mode respectively. Similarly as for the single-mode method, engineering a correlated dissipator with an ancillary qubit mode of the form $\mathcal{D}[\L_{CX} \sp]$ would achieve a discrete CNOT gate, up to exponentially small corrections. However, the projection operators ${P}_\pm$ are highly non-local, and therefore this scheme is of questionable use. First, such operators are not currently implementable with superconducting circuits, although there is a path towards them as high-impedance operators~\cite{cohen2017autonomous}. Second, if one assumes access to such ``next-generation'' operators, then for fairness one should view as relatively easy too the options of e.g.~Section \ref{sec:flathamil}. Third, engineering non-local operators in a cat-qubit context is a Pandora's box, as it introduces artificial operators against which the exponential bit-flip protection is \textit{not} designed to work.

Other designs may take inspiration from the Zeno implementation of the CNOT gate, i.e.~inducing, in superposition, $n-n_p$ discrete $Z(\theta)$ gates on the control qubit conditional on Fock state $n$ in the target qubit. In the Zeno Hamiltonian implementation, this indeed remains compatible with exponential bit-flip protection. Generalizing this to a jump operator appears to require a complicated ancilla system. For instance, if we would rely on $n-n_p$ ancilla jumps to perform the right number of $Z(\theta)$ gates, then Fock-number information would leak out to the environment unless specific erasure actions are taken to impeach detecting the number of jumps. In absence of concrete insight on realistic experimental building blocks associated to such ancilla, we leave this for future research.

Barring the experimental difficulty of realizing an operator like $\L_{CX}$, the effect of unwanted $\ket{g} \leftrightarrow \ket{e}$ transitions of the ancillary mode that monitors the gate warrants a caveat. Indeed, an unwanted transition would mean applying the CNOT gate possibly an even number of times, which would by definition involve both phase-flip and bit-flip errors. In the traditional model, bit-flip errors remain exponentially suppressed even for CNOT gates, thanks to a smart use of the available continuous phase space which we lose with this design. On the upside, the particular use of the ancillary qubit in this discrete gate opens the door to specific designs reducing ancilla-induced errors, as discussed above for the $Z(\theta)$ gate.


\section{Conclusion}\label{sec:conclusions}
We have introduced four new designs of dissipative cat qubit gates that can help reduce gate-induced phase errors, and therefore help reach error-correction thresholds. Upon the observation that incoherent gate errors result from the entanglement between the cat qubit and its buffer mode, we have devised two designs meant to take advantage of the buffer memory in the system. The first one is based on the photodetection of the buffer mode output, and thus on the retrieval of information that can then be classically fed back. The precision of this scheme is only limited by the photodetector efficiency. The second one relies on an autonomous error correcting scheme for which first-order gate-induced parity errors are automatically corrected for as buffer photons exit the resonator. In particular, this second design is readily implementable with superconducting circuits and achieves a reduction of up to two orders of magnitude in gate errors. The same setup can also be used with squeezed cat states for the autonomous correction of single-photon losses~\cite{xu2022autonomous}, currently one of the main limiting factors of bosonic qubits.

We have also described two drastically different gate designs. By engineering higher-order drive Hamiltonians that feature locally flat energy potentials in position representation, spurious effects on the cat qubit mode can be avoided and an improvement in gate error fidelities is achieved. We believe that such Hamiltonians could also be used for other purposes, such as cat state preparation. Finally, we have explored how to engineer cat qubit gates without the Zeno effect, and introduced dissipation-based gate engineering. By coupling the cat mode to an ancillary nonlinear mode that monitors the ongoing gate, a discrete $\pi$-phase gate is realized. This method can circumvent the usual linear gate time scaling of Zeno-based gates and achieve exponentially scaling gate fidelities, and associated CNOT gate designs are therefore our subject of ongoing investigation.

We hope that this article has been able to provide its readers intuition about the design of cat qubit gates. While we have explored many paths towards high-fidelity gates, we believe that further improvements can still be achieved, in particular for multi-mode encodings of cat qubits. Finally, while these gate designs have been particularly focused on dissipative cat qubits, they should also inspire the design of superadiabatic operations on any bosonic or dissipatively-stabilized system. In particular, most tailored dissipation operators often rely on a highly damped buffer mode together with adiabatic elimination. We have thus described for the first time how the dynamics of this buffer mode can be taken advantage of.

\begin{acknowledgements}
The authors thank Benjamin Huard, Pierre Rouchon, Lev-Arcady Sellem and Jérémie Guillaud for fruitful discussions. This work was supported by the French Agence Nationale de la Recherche under grant ANR-18-CE47-0005. We also acknowledge funding from the Plan France 2030 through the project ANR-22-PETQ-0006. The numerical simulations were performed using HPC resources at Inria Paris, and using the QuTIP open-source package. The authors thank its developers and maintainers for their work.
\end{acknowledgements}


\bibliography{bibliography}


\clearpage
\appendix
\renewcommand{\appendixname}{APPENDIX}


\section{REVIEW OF THE SHIFTED FOCK BASIS}\label{sec-apdx:gatereview}

In this appendix, we review the Shifted Fock Basis (SFB) as first introduced in~\cite{chamberland2020building}. We refer to the original paper for a full review to the method, but this short section provides the minimal elements required for the comprehension of the main text.

The SFB is an alternative basis of states for a quantum oscillator which splits the main properties of a cat qubit in two separate modes: a qubit mode that represents the logical cat qubit state, and a gauge mode that represents leakage away from the stabilized computational subspace~\eqref{eq:logicalstates}. More precisely, let us define the non-orthonormal basis states
\begin{equation}\label{eq:sfbstates}
    \ket{\pm, n} \equiv \frac{1}{\sqrt{2}} \big[ \vb*{D}(\alpha) \pm (-1)^n \vb*{D}(-\alpha)\big] \ket{\vb*{n} = n}
\end{equation}
where $\vb*{D}(\alpha)$ is the displacement operator and $\ket{\vb*{n} = n}$ is the $n$-th Fock state. These states are such that, in the ground state manifold $n=0$, they match the cat qubit computational basis states, \ie~$\ket{\pm, 0} \propto \ket{\pm}_L$. Although these states do not define an orthonormal basis, any two states in a separate $\pm$ branch are exactly orthogonal due to the even/odd parity in the number of photons. In addition, the first shifted Fock states are approximately orthonormal in the limit of $|\alpha| \gg 1$. To rigorously treat this basis, it can be orthornormalized for instance with a Gram-Schmidt process, but we again refer to~\cite{chamberland2020building} for details.

A particularly interesting property of the SFB is how it transforms the annihilation operator $\a$. Indeed, acting on the basis states of~\eqref{eq:sfbstates} yields
\begin{equation}\label{eq:sfbannihilation}
    \begin{split}
        \a \ket{\pm, n} &= \sqrt{n} \ket{\mp, n-1} + \alpha \ket{\mp, n} \\
        &= \sz \otimes (\ta + \alpha) \ket{\pm, n}
    \end{split}
\end{equation}
where we have defined two new operators $\sz$ and $\ta$ such that $\sz \ket{\pm, n} = \ket{\mp, n}$ and $\ta \ket{\pm, n} = \sqrt{n} \ket{\pm, n-1}$ in analogy to a qubit and quantum oscillator mode. From~\eqref{eq:sfbannihilation} we infer the mapping of the annihilation operator from the Fock to the shifted Fock basis, which reads
\begin{equation}
    \a \rightarrow \sz \otimes (\ta + \alpha).
\end{equation}
Again, this mapping is only valid in the limit of small shifted Fock excitation numbers because of the non-orthonormalization of the basis. It is however very helpful for understanding the intimate dynamics of cat qubit gates and for the numerical analysis of phase-flip errors.

In particular, it is possible to derive the probability of phase errors induced by cat qubit gates up to first-order in the dynamics. This derivation was done extensively in~\cite{chamberland2020building}, so we only give these results again here for completeness. For single-qubit $Z(\theta)$ gates, the probability of non-adiabatic phase errors reads
\begin{equation}\label{eq:zgateerr}
    p_Z = \frac{\theta^2}{16 |\alpha|^4 \kappa_2 T}
\end{equation}
where $\kappa_2 \equiv 4 g_2^2 / \kappa_b$ in the presence of two-photon exchange coupling with a buffer mode. For two-qubit CNOT gates, non-adiabatic phase errors only affect control qubits, and read
\begin{equation}\label{eq:cnot-errs}
    p_{Z_C} = \frac{\pi^2}{16 |\alpha|^2 \kappa_2 T}, \quad p_{Z_T} = p_{Z_T Z_C} = 0
\end{equation}
where the subscripts $C$ and $T$ stand for control and target qubits respectively.


\section{EFFECT OF NOISE AND SPURIOUS HAMILTONIANS}\label{sec-apdx:errormodel}

In the main text, we study gate designs with simplified master equation models in order to capture the key elements of each proposal. However, actual experimental setups feature various sources of errors that could in principle hinder gate performances. In this appendix, we study a more complete model of errors to demonstrate that our proposals hold even under realistic noise processes.

The master equation we consider is
\begin{equation}
    \begin{split}
        \frac{d \vrho}{dt} &= \mathcal{L}_{0} \vrho - i \left[ \H_s, \vrho \right] + \kappa_a (1 + n_{th, a}) \mathcal{D}[\a] \vrho \\
        &+ \kappa_a n_{th, a} \mathcal{D}[\adag] \vrho + \kappa_{\phi, a} \mathcal{D}[\adag \a] \vrho
    \end{split}
\end{equation}
where $\mathcal{L}_0$ denotes the Liouvillian to be engineered on the system --- varied from one gate design to the next --- which typically contains two-to-one photon exchange, buffer mode dissipation and a gate drive on the memory mode, and
\begin{equation}
    \H_s = -K_a \a^{\dagger 2} \a^2 + \chi_{ab} \adag \a \bdag \b - K_b \b^{\dagger 2} \b^2
\end{equation}
is a spurious Hamiltonian with Kerr and cross-Kerr terms on the memory and buffer modes. These terms typically result from the residual cosine potential in the full circuit Hamiltonian, but in practice, they can be small compared to two-photon dissipation when the system is engineered with an Asymmetrically Threaded SQUID at the appropriate flux bias point~\cite{lescanne2020exponential}. Our master equation model also includes single-photon losses, thermal photons and pure dephasing on the memory mode with respective rates $\kappa_a$, $n_{th, a}$ and $\kappa_{\phi, a}$. Note that we do not include thermal photons on the buffer mode in this model as they are discussed separately in Appendix~\ref{sec-apdx:thermalbuffer}.

\begin{figure*}[!t]
         \centering 
         \includegraphics[width = \linewidth]{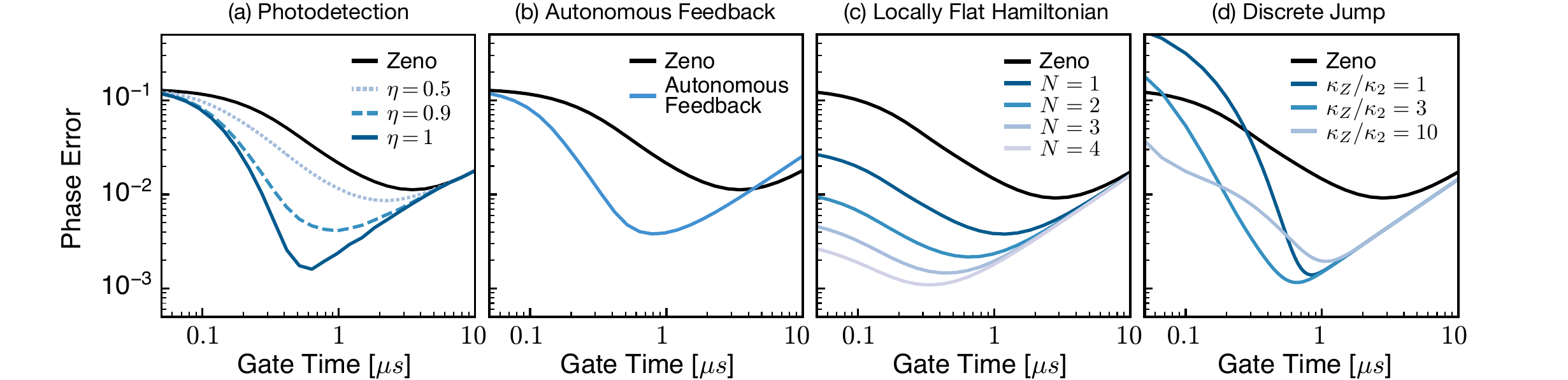}
    \label{fig:errormodel}
    \vspace{-0.5cm}
    \caption{
        Phase errors during a $Z(\pi)$ gate under a complete model of noise for each gate design introduced in the main text. Each design is compared to the standard Zeno gate (black). In all panels, $|\alpha|^2 = 4$ and energies are set to $g_2 / 2\pi = 1\,\mathrm{MHz}$, $K_a / 2\pi = 1\,\mathrm{kHz}$, $K_b / 2\pi = 810\,\mathrm{kHz}$, $\chi_{ab} / 2\pi = 65\,\mathrm{kHz}$, $\kappa_a / 2\pi = 53\,\mathrm{Hz}$, $n_{th, a} = 10\%$ and $\kappa_{\phi, a} / 2\pi = 10\,\mathrm{Hz}$. For panels (a), (c) and (d), $\kappa_b / 2\pi = 8\,\mathrm{MHz}$. For panel (b), $\kappa_{ab} / 2\pi = 2\,\mathrm{MHz}$.
    }
 \end{figure*}

To study these noise processes, we consider specific values for each term mainly extracted from Refs.~\cite{lescanne2020exponential,reglade2023quantum}. The two-to-one photon exchange coupling is $g_2 / 2\pi = 1\,\mathrm{MHz}$ and combined with a buffer mode dissipation at $\kappa_b / 2\pi = 8\,\mathrm{MHz}$ which yields an effective two-photon dissipation rate $\kappa_2 / 2\pi = 500\,\mathrm{kHz}$. Kerr and cross-Kerr energies are $K_a / 2\pi = 1\,\mathrm{kHz}$, $K_b / 2\pi = 810\,\mathrm{kHz}$ and $\chi_{ab} / 2\pi = 65\,\mathrm{kHz}$. Single-photon loss is $\kappa_a / 2\pi = 53\,\mathrm{Hz}$ corresponding to $T_1 = 3\,\mathrm{ms}$ in the memory mode, which is typical in current experiments with high-Q superconducting resonators. Note that lower cavity lifetimes could also be investigated, but would limit the potential gain in gate fidelity of our proposals compared to standard Zeno gates as phase errors would be dominated by single-photon losses in this regime. Finally, thermal noise and pure dephasing are $n_{th, a} = 10\%$ and $\kappa_{\phi, a} / 2\pi = 10\,\mathrm{Hz}$.

In Figure~\ref{fig:errormodel}, gate errors under this noise model are shown for each gate design from the main text. Each time, the performance is compared to the standard Zeno gate for a $Z(\pi)$ gate, shown in black lines. In all panels, phase errors eventually converge to a linear dependence in the large gate time regime in which single-photon losses dominate. In addition, there is systematically an optimal gate time which minimizes gate errors and that represents the optimal trade-off between gate-induced and single-photon loss errors.

For the photodetection design in panel (a), we vary the photodetection efficiency from a perfect photodetector $\eta = 1$ to a non-ideal one at $\eta = 0.5$, demonstrating again the linear dependence with detection inefficiency as discussed in the main text. Panel (b) shows the performance of the autonomous feedback design with $\kappa_{ab} / 2\pi = 2\,\mathrm{MHz}$ and $\kappa_b = 0$. In panel (c), we show the locally flat Hamiltonian design for increasing number of odd drive terms from $N=1$ to $N=4$ each time gaining in optimal gate fidelity and gate time. Finally, panel (d) shows performances of the discrete jump design where the optimal gate fidelity depends on the ratio of $\kappa_Z$ to $\kappa_2$ and on the actual error model studied.


\section{THERMAL NOISE IN THE BUFFER MODE}\label{sec-apdx:thermalbuffer}

In this appendix, we discuss the role of thermal noise in the buffer mode for the first two designs based on the feedback of information introduced in the main text. Since these two schemes use the buffer mode population to detect potential phase-flips induced on the memory mode, it is quite natural that one of their main limitations should come from spurious population in the buffer.

\subsection{Photodetection}

For the photodetection scheme of Section~\ref{sec:photodetection}, the feedback action to apply is a $Z(\pi)$ gate after every buffer photon detection. Therefore, the rate of thermal photons is directly linked to the rate of phase information loss on the cat state, assuming that the feedback action is perfect. From the point of view of the Shifted Fock Basis, one can adiabatically eliminate the gauge mode in~\eqref{eq:sme-feedback} and average out the stochastic terms. This leads to a simplified model in the absence of a gate drive,
\begin{equation}
    \begin{split}
        \frac{d\vrho}{dt} &= \kappa_b \eta (1 + n_{th, b}) \mathcal{D}[\b \sz] \vrho \\
        &+ \kappa_b (1 - \eta) (1 + n_{th, b}) \mathcal{D}[\b] \vrho \\
        &+ \kappa_b n_{th, b} \mathcal{D}[\bdag] \vrho
    \end{split}
\end{equation}
Here, the beamsplitter interaction between gauge and buffer modes is eliminated compared to the fast dynamics on the buffer. In particular, notice the correlated dissipation in $\mathcal{D}[\b \sz]$ which arises from the idealized classical feedback action on the memory mode. From this master equation, we easily compute the average expectation value of the $\sx$ operator by going to the Heisenberg picture, which reads
\begin{equation}
    \langle \sx \rangle(t) = \langle \sx \rangle(t=0) \times \exp(-\gamma t)
\end{equation}
where $\gamma = 2\eta \kappa_b n_{th, b} (1+n_{th,b})$ rate of parity information loss in the presence of thermal noise in the buffer. Note that the classical feedback action is always optional, and typically it should only be turned on during gates. To know whether the gate design is beneficial in practice, this rate should be compared to the rate of single-photon losses $\kappa_a$ which is the usual dominating source of loss of parity information.

\begin{figure}[!t]
         \centering 
         \includegraphics[width = \columnwidth]{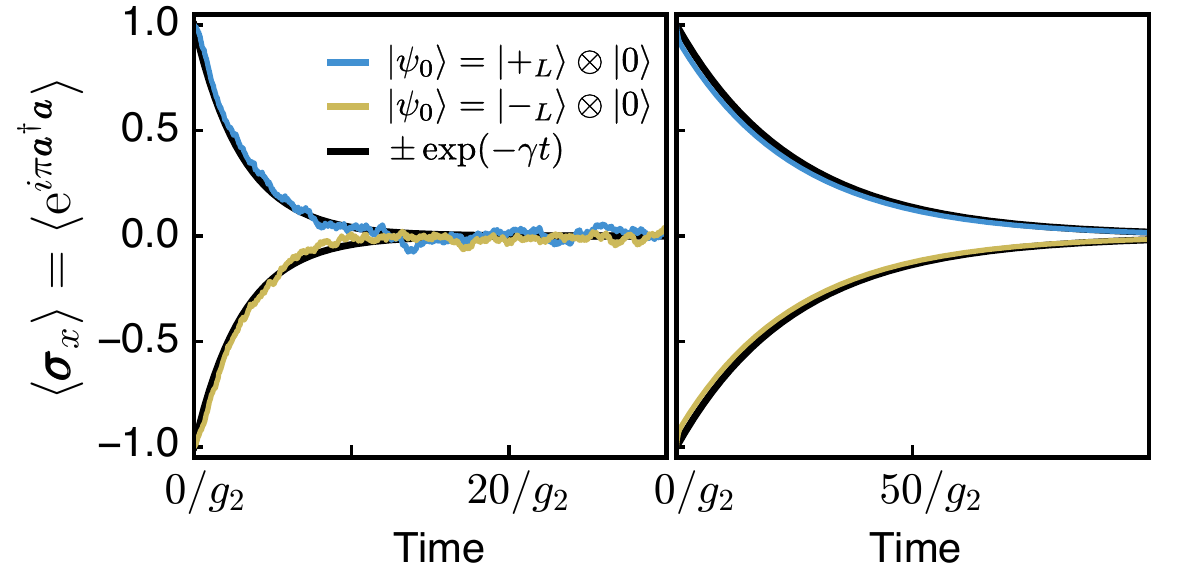}
    \label{fig:thermalbuffer}
    \vspace{-0.5cm}
    \caption{
        (Left) Photon-number parity of an idling cat state under the photodetection and classical feedback design of Sec.~\ref{sec:photodetection} with thermal noise in the buffer mode $n_{th, b} = 2\%$. Parity is averaged over 1000 stochastic trajectories, and fitted with an exponential decay at rate $\gamma = 2 \eta \kappa_b n_{th, b} (1 + n_{th, b})$. In this simulation, $\kappa_b / g_2 = 8$, $|\alpha|^2 = 8$ and $\eta=1$. (Right) Photon-number parity of an idling cat state under the autonomous feedback of Sec.~\ref{sec:autonfeedback} with thermal noise in the reservoir mode $n_{th, r} = 2\%$. Parity is fitted with an exponential decay at rate $\gamma = 2 \sqrt{g_2 \kappa_{ab}} n_{th, r}$. In this simulation, $\kappa_{ab} |\alpha|^2 / g_2 = 8$ and $|\alpha|^2 = 8$.
    }
 \end{figure}

On the left panel of Figure~\ref{fig:thermalbuffer}, we show the parity of an idling cat state initialized in $\ket{\pm_L}$ as obtained from numerical integration of the full stochastic master equation (i.e. including two-photon coupling, buffer mode dissipation and thermal noise in the buffer). The solution is averaged over 1000 stochastic trajectories, and we find a very good fit with the simplified model as described above.

\subsection{Autonomous feedback}

For the feedback design of Section~\ref{sec:autonfeedback}, the feedback action is this time performed autonomously through a correlated dissipation operator in $\mathcal{D}[\a\b]$. Buffer excitations would then also result in spurious parity swaps on the cat mode. However, with this design, the buffer mode is \textit{a priori} a high-Q mode since the correlated dissipation is engineered through a third reservoir mode $\r$ as discussed in the main text, with a three mode coupling of the form $g (\a \b \rdag + \hc)$. Therefore, the main limitation of this design does not come from buffer mode excitations, but rather from thermal noise in the reservoir mode in the form $\kappa_r n_{th, r} \mathcal{D}[\rdag]$. From adiabatic elimination of this reservoir with the formalism of~\cite{azouit2016adiabatic,forni2018adiabatic}, we derive an effective rate of correlated excitations. The effective master equation on the memory and buffer modes for an idling cat qubit thus reads
\begin{equation}\label{eq:mastereq-thermal-auton}
    \begin{split}
        \frac{d \vrho}{dt} &= -i [\H_{AB}, \vrho] \\
        &+ \kappa_{ab} (1 + n_{th, r}) \mathcal{D}[\a\b] \vrho \\
        &+ \kappa_{ab} n_{th, r} \mathcal{D}[\adag \bdag] \vrho
    \end{split}
\end{equation}
where a new term in $\mathcal{D}[\adag \bdag]$ appears compared to the ideal master equation. This dissipation creates correlated excitations on the memory and buffer modes. If these correlated excitations are dissipated through the $\mathcal{D}[\a\b]$ term, then the parity is swapped back to its original value and no phase information is lost. However, the correlated excitation can also undergo partial two-photon Rabi oscillations through the $\H_{AB}$ term and then eventually dissipate away. In this case, the parity is not strictly preserved by the dynamics and phase information is lost to the environment.

On the right panel of Figure~\ref{fig:thermalbuffer}, we show the parity of an idling cat state initialized in $\ket{\pm_L}$ as obtained from the numerical integration of~\ref{eq:mastereq-thermal-auton}. We indeed find that the parity information decays exponentially with a rate $\gamma$. For all values of $g_2$, $\kappa_{ab}$ and $n_{th, r}$ we investigated, this rate seems to fit the formula $\gamma = 2 \sqrt{g_2 \kappa_{ab}} n_{th, r}$. In particular, turning off either two-to-one photon coupling or thermal noise on the reservoir mode suppresses this effect.


\section{MASTER EQUATION INVARIANCE BY JOINT PHASE CONJUGATION}\label{sec-apdx:xaxisconjugation}

In this appendix, we show how the Lindblad master equation of a $Z(\theta)$ gate is invariant under a joint x-axis phase conjugation of the cat and buffer modes. 

\subsection{Phase conjugation superoperator}

An x-axis phase conjugation consists in flipping the sign of quadrature $\x$ while keeping quadrature $\p$ unchanged~\cite{cerf2001phase}. While this is an unphysical transformation --- it does not preserve commutation relations ---, it can be understood as a time-reversal operator. It is also a standard example of a superoperator which is positive but not completely positive~\cite{wolf2008dividing}. In our case, the symmetry of the master equation by joint phase conjugation implies that the logical bit information of the cat qubit (encoded in the x-axis) is also encoded in the x-axis of the buffer mode, and so in an exact manner. 

Let us first define the x-axis phase conjugation superoperator on a single mode, which we denote as $\mathcal{C}$. By definition, this superoperator is such that $\mathcal{C} \ketbra{x}{x} = \ketbra{-x}{-x}$ and $\mathcal{C} \ketbra{p}{p} = \ketbra{p}{p}$ where $\ket{x}$ and $\ket{p}$ are quadrature eigenstates. By linearity, the identities $\mathcal{C}\x = -\x$ and $\mathcal{C}\p = \p$ directly follow since $\x = \int d x \cdot x \ketbra{x}{x}$ and $\p = \int d p \cdot p \ketbra{p}{p}$. More generally, we have that $\mathcal{C} f(\x) = f(-\x)$ for any function $f$. Reinserting $\ket{x} = \int d p \, \e^{-ipx} \ket{p}$ in this identity, we have that
\begin{equation}\label{eq:conjugderiv}
    \begin{split}
        & \int d x f(x) \iint d p d p' \e^{-i (p - p') x} \cdot  \mathcal{C}\ketbra{p}{p'} \\
        & = \int d x f(x) \iint d p d p' \e^{i (p - p') x} \cdot \ketbra{p}{p'}
    \end{split}
\end{equation}
Rearranging the integrals and denoting $\tilde{f}(p)$ the Fourier transform of $f(x)$,~\eqref{eq:conjugderiv} reads
\begin{equation}
    \iint d p d p' \tilde{f}(p-p')  \cdot  \left(\mathcal{C}\ketbra{p}{p'} - \ketbra{p'}{p} \right) = 0
\end{equation}
Since this equation holds for any function $\tilde{f}$, then we have that $\mathcal{C} \ketbra{p}{p'} = \ketbra{p'}{p}$, and similarly for the other quadrature, $\mathcal{C} \ketbra{x}{x'} = \ketbra{-x'}{-x}$. In particular, notice the transposition in both of these relations, which accounts for the time-reversal property of the conjugation. 

Before moving on, there is a last identity that will be useful for our upcoming derivation. For any density matrix $\vrho$, we have
\begin{equation}\label{eq:conjugidentity}
    \begin{split}
        \mathcal{C}(\x \vrho) &= \mathcal{C} \left( \iint d x d x' x \rho(x, x') \ketbra{x}{x'} \right) \\ 
        &= \iint d x d x' x \rho(x, x') \ketbra{-x'}{-x} \\
        &= \iint d x' d x (-x') \rho(-x', -x) \ketbra{x}{x'} \\
        &= - \mathcal{C}(\vrho) \x \, .
    \end{split}
\end{equation}
Using that $\mathcal{C}^{2}$ yields the identity superoperator, we directly get the converse identity, $\mathcal{C}(\vrho \x) = - \x \mathcal{C}(\vrho)$. Similarly for the other quadrature, we have that $\mathcal{C}(\p \vrho) = \mathcal{C}(\vrho) \p$ and $\mathcal{C}(\vrho \p) = \p \mathcal{C}(\vrho)$.

\subsection{Master equation invariance}
Our goal is now to show that the joint phase conjugation of the cat and buffer modes is a symmetry of the $Z(\theta)$ gate dynamics. This involves showing that the joint phase conjugation superoperator $\mathcal{C}_{AB} = \mathcal{C}_A \otimes  \mathcal{C}_B$ commutes with the Lindblad superoperator $\mathcal{L}$, where
\begin{equation}
    \mathcal{L} = \mathcal{L}_Z + \mathcal{L}_{AB} + \mathcal{L}_{1,A} + \mathcal{L}_{1,B}
\end{equation}
where $\mathcal{L}_Z = -i[\H_Z, \cdot \; ]$, $\mathcal{L}_{AB} = -i[\H_{AB}, \cdot \; ]$, $\mathcal{L}_{1,A} = \kappa_a \mathcal{D}[\a]$ and $\mathcal{L}_{1,B} = \kappa_b \mathcal{D}[\b]$. By linearity, it is sufficient to show the commutation relation for each term in the Lindblad superoperator. For the single-mode cavity drive with Hamiltonian $\H_Z = \varepsilon_Z \x_a$, we get
\begin{equation}
    \begin{split}
        \mathcal{C} (\mathcal{L}_Z (\vrho)) &= -i \varepsilon_Z \mathcal{C} (\x_a \vrho - \vrho \x_a) \\
        &= -i \varepsilon_Z \left(-\mathcal{C}(\vrho) \x_a + \x_a \mathcal{C}(\vrho)\right) \\
        &= \mathcal{L}_Z(\mathcal{C}(\vrho)) \, .
    \end{split}
\end{equation}
For the two-photon exchange term, the Hamiltonian reads $\H_{AB} = g_2 (\a^2 - \alpha^2) \bdag + \hc$. Rewriting the Hamiltonian in terms of quadrature operators yields
\begin{equation}
    \H_{AB} / g_2 = (\x_a^2 - \p_a^2) \frac{\x_b}{4} + (\p_a \x_a + \x_a \p_a) \frac{\p_b}{4} - \alpha^2 \x_b \, .
\end{equation}
Since this Hamiltonian features only terms with an odd number of quadrature $\x$ and even number of quadrature $\p$ operators, using the identities~\eqref{eq:conjugidentity} and the related ones yields an overall minus sign and a transposition, from which we infer
\begin{equation}
    \mathcal{C}(\mathcal{L}_{AB}(\vrho)) = \mathcal{L}_{AB}(\mathcal{C}(\vrho)) \, .
\end{equation}
Finally, for the single-photon loss operators on either mode $\mathcal{L}_1 = \mathcal{D}[\x + i\p]$, and again using the~\eqref{eq:conjugidentity} identities, 
\begin{equation}
    \begin{split}
        \mathcal{C} \big(\mathcal{L}_1(\vrho) \big) &= \mathcal{C}\big( (\x + i\p) \cdot \vrho \cdot (\x - i\p)\big) \\ 
        & \qquad - \mathcal{C} \big( (\x^2 + \p^2) \cdot \vrho \big) / 2 \\
        & \qquad - \mathcal{C} \big( \vrho \cdot (\x^2 + \p^2) \big) / 2 \\
        &= (\x + i\p) \cdot \mathcal{C}(\vrho) \cdot (\x - i\p) \\ 
        & \qquad - \mathcal{C}(\vrho) \cdot (\x^2 + \p^2) / 2 \\
        & \qquad - (\x^2 + \p^2) \cdot \mathcal{C}(\vrho) / 2 \\
        &= \mathcal{L}_1 (\mathcal{C}(\vrho))
    \end{split}
\end{equation}
By linearity, the required result is shown, $\mathcal{C} \mathcal{L} - \mathcal{L} \mathcal{C} = 0$. It shows the invariance of the $Z(\theta)$ gate master equation under a joint x-axis phase conjugation of both cat and buffer modes.


\section{IDEAL FEEDBACK FOR PHOTODETECTION}\label{sec-apdx:feedback}

In this appendix, we discuss the ideal feedback action to be applied for the photodetection design of~\ref{sec:photodetection}. Assuming a perfect photodetector with detection efficiency $\eta = 1$ and instantaneous feedback, this ideal feedback can yield error-less gates. Indeed, if $\eta = 1$, then no information is lost to the environment and the combined qubit-buffer state remains pure at all times, such that the final cat-qubit state is pure. An appropriate feedback action can thus retrieve the desired angle of rotation.

\begin{figure}[!t]
         \centering
         \includegraphics[width = \columnwidth]{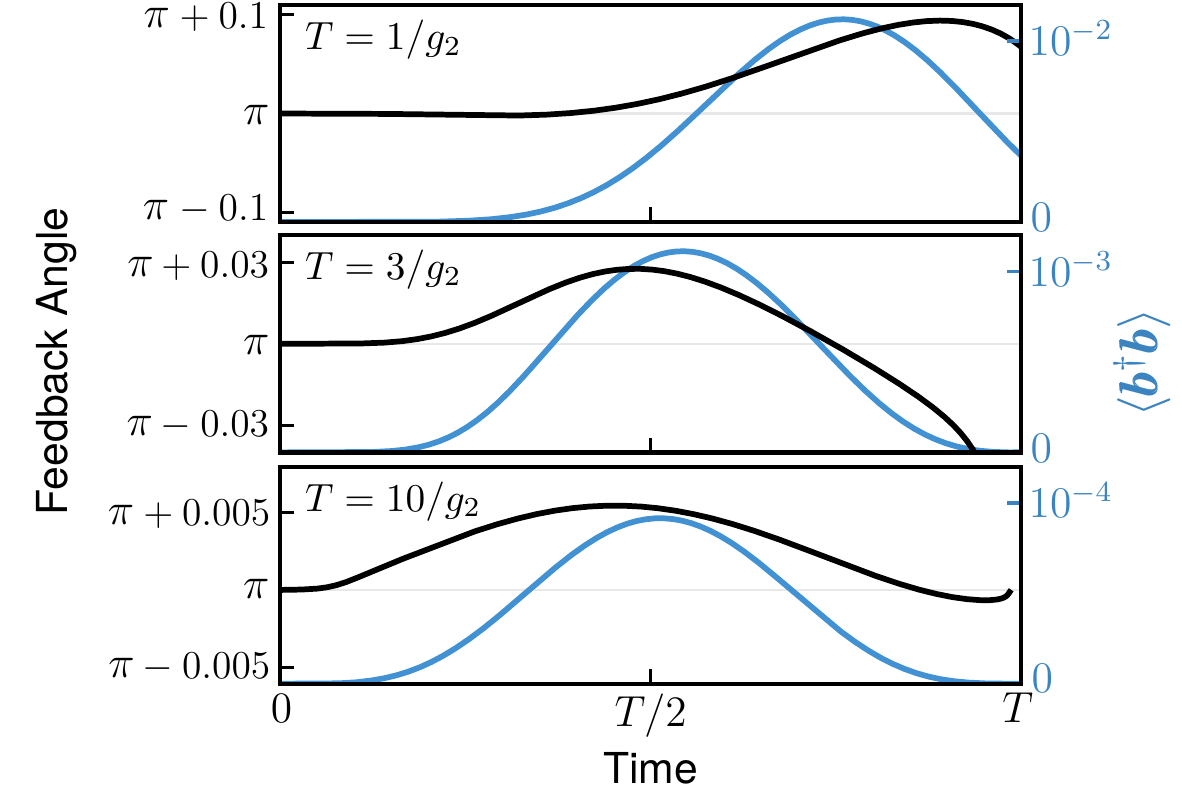}
    \label{fig:PD-jumpangle}
    \vspace{-0.5cm}
    \caption{
        (Black) Angle of the $Z$ rotation feedback to perform after a buffer mode photodetection at any given time, for a $Z(\pi)$ gate of duration $T$. (Blue) Buffer mode population for the no-jump trajectory, proportional to the photodetection probability $\langle d N_\eta \rangle$. In these numerical simulations, $\kappa_b = 8 g_2$ and $|\alpha|^2 = 8$.
    }
 \end{figure}

Figure~\ref{fig:PD-jumpangle} shows the profiles of optimal feedback angles $\pi+\delta(t_J)$ upon photon detection, as a function of detection time $t_J$, for a targeted $Z(\pi)$ gate. The mean photon population of the buffer mode is also shown as a function of time, which is proportional to the instantaneous photodetection probability $\langle d N_\eta \rangle$. Note that close to $t=T$ in the middle plot, the angle $\delta$ diverges to negative values, but $\langle \bdag \b \rangle \ll 1$ at this time such that it is excessively unlikely to ever have to perform this feedback action. 

To perform this classical feedback upon detection of buffer photons, multiple solutions exist. The first and most straightforward is to actually add this Z($\pi+\delta$) gate to the actuation, which would \textit{occasionally} increase the gate time depending on the input gate angle $\theta$ and detection time $t_J$. For instance, if a photon is detected at the beginning of a $Z(0.9\pi)$ gate, then a shorter $Z(-0.1\pi+\delta)$ gate should be performed instead, while if it is detected towards the end then a longer $Z(1.9\pi+\delta)$ gate would be realized. While this happens only for photon-detected trajectories, and thus with low probability, it causes unpredictable gate times. 

As a second option, the feedback could adjust the final $Z(\theta)$ gate up to an integer multiple of $\pi$ (similarly, adjust the CNOT gate up to an integer number of $Z(\pi)$ on the control qubit). The remaining gate $Z(n\pi)$ is a Pauli gate, which either a specific hardware operation can implement or the software can keep track of by adapting remaining computer operations until the next non-Clifford gate is reached.

For CNOT gates, the output of the target buffer mode should in principle also be monitored due to the reconvergence phase in which the target mode undergoes significant dynamics. Importantly, the effect of this monitoring --- whatever the detection results --- is only on the phase of the control qubit. Indeed, the bit degrees of freedom remain exponentially protected on both qubits; and since no operator ever couples to the bit-value information of the target qubit (unlike the control qubit, for which $\a+\adag$ is sensitive to the bit value), it means that its bit-value information can never leak out in this perfectly monitored scheme, and hence no phase-blurring backaction can happen on the target qubit. We have indeed verified numerically that, for $\eta=1$ and no buffer photons detected, an ideal CNOT gate is retrieved after an infinite time reconvergence by adjusting a $Z(\delta)$ gate on the control qubit with $\delta \ll 1$. A detection on the control qubit buffer output at time $t_J$ requires an additional $Z(\pi+\delta(t_J))$ gate, as for the previously discussed $Z(\theta)$ gate scheme. In contrast, a detection event on the output of the target buffer mode still requires corrections of order $\delta$ only. This is consistent with a first-order Heisenberg picture analysis as in Section \ref{sec:gateerrors}, for which $\b_{C}$ is proportional to $\vb*{\sigma}_{z,C}$ while $\b_{T}$ does not carry any qubit information.


\section{ADIABATIC ELIMINATION OF THE BUFFER MODE}\label{sec-apdx:adiabelim}

In Section~\ref{sec:autonfeedback} of the main text, the adiabatic elimination of the buffer mode is performed in the presence of a correlated dissipator. The two-mode master equation is initially given by
\begin{equation}
    \frac{d \vrho}{dt} = -i \left[ \H_{AB}, \vrho \right] + \kappa_{ab} \mathcal{D}[\a\b] \vrho.
\end{equation}
This equation describes an exchange term between cat and buffer modes, as well as a strong correlated dissipation. Here, the fast dynamics corresponds to the deexcitation of the buffer mode thanks to the dissipation term, while the slow dynamics is that of the exchange Hamiltonian. The goal is thus to adiabatically eliminate the fast dynamics when the buffer is excited, and to derive an effective single-mode equation for the A mode. Taking the notations of~\cite{reiter2012effective}, we have
\begin{equation}
    \begin{split}
        \vb*{V}_+ &= g_2 (\a^2 - \alpha^2) \bdag \\
        \vb*{V}_- &= g_2^* (\a^{\dagger 2} - \alpha^{* 2}) \b
    \end{split}
\end{equation}
which are perturbative (de-)excitations of the system, $\vb*{L} = \sqrt{\kappa_{ab}} \a \b$ is the jump operator from excited to ground subspaces, and $\H_g = \H_e = 0$ are the block diagonal Hamiltonians in the ground and excited subspaces. In addition, the non-hermitian Hamiltonian in the excited subspace reads
\begin{equation}
    \H_\text{NH} = -\frac{i}{2} \kappa_{ab} \adag \a \bdag \b \, .
\end{equation}
Therefore, the effective single-mode dynamics reads
\begin{equation}
    \frac{d \vrho}{dt} = -i [\H_\text{eff}, \vrho] + \mathcal{D}[\vb*{L}_\text{eff}] \vrho
\end{equation}
where $\H_\text{eff} \propto \H_\text{NH}^{-1} + \H_\text{NH}^{\dagger -1} = 0$, and
\begin{equation}
    \begin{split}
        \vb*{L}_\text{eff} &= \vb*{L} \left(\H_\text{NH}\right)^{-1} \vb*{V}_+ \\
        &= \frac{2 i g_2}{\sqrt{\kappa_{ab}}} \a (\adag \a)^{-1} (\a^2 - \alpha^2)
    \end{split}
\end{equation}
This describes a parity-switching jump operator with cat qubit steady states. In the semi-classical limit, $\a (\adag \a)^{-1} \sim \alpha^{-1}$, such that the effective two-photon dissipation rate is indeed given by $\kappa_2 \equiv 4 g_2^2 / \alpha^2 \kappa_{ab}$.


\section{OPTIMIZING THE FLATNESS OF DRIVE HAMILTONIANS}\label{sec-apdx:polyn}

In Section~\ref{sec:flathamil} of the main text, we define drive Hamiltonians with odd-power polynomials of the position operator $\x$. For cat qubit gate engineering, it is required that these high-order drives are locally flat around both cat qubit coherent components. Mathematically, this corresponds to a minimization of the variance over one coherent state, defined by
\begin{equation}\label{eq:polyn-apdx-var}
    V_N\left(\{c_n \}\right) \equiv \frac{1}{\sqrt{2\pi}}\int_{-\infty}^\infty H_{Z,N}(x)^2 \e^{-\frac{1}{2}\left(x-2\alpha\right)^2} d x ,
\end{equation}
under the constraint of a fixed mean value of the drive over this coherent state, defined by
\begin{equation}\label{eq:polyn-apdx-mean}
    E_N\left(\{c_n \}\right) \equiv \frac{1}{\sqrt{2\pi}}\int_{-\infty}^\infty H_{Z,N}(x) \e^{-\frac{1}{2}\left(x-2\alpha\right)^2} d x
\end{equation}
where $H_{Z,N}(x) = \varepsilon_Z \sum_{n=0}^N c_n x^{2n+1}$ is the potential to be optimized, with $N+1$ constants to be determined.

To perform this constrained minimization problem, we use Lagrange multipliers, and define the Lagrangian function as
\begin{equation}\label{eq:polyn-apdx-lagrange}
    \mathcal{L}_N\left(\{ c_n \}, \lambda\right) = V_N\left(\{c_n \}\right) - \lambda \left(E_N\left(\{c_n \}\right) - \varepsilon_0\right)
\end{equation}
where $\lambda$ is a Lagrange multiplier, and $\varepsilon_0$ is the fixed mean value of $E_N$. 

Thanks to the simple form of~\eqref{eq:polyn-apdx-var} and~\eqref{eq:polyn-apdx-mean}, it is quite simple to find the global minimum of this Lagrangian function exactly. Differentiating~\eqref{eq:polyn-apdx-lagrange} with respect to all $N+2$ variables yields
\begin{equation}
    \left\{
    \begin{aligned}
        &~\frac{\partial \mathcal{L}_N}{\partial c_k} = 2 \sum_{n=0}^N I_{2(n+k+1)} c_n - \lambda I_{2k+1} \\
        &~\frac{\partial \mathcal{L}_N}{\partial \lambda} = 1 -  \sum_{n=0}^N I_{2n+1} c_n
    \end{aligned}
    \right.
\end{equation}
where we have defined 
\begin{equation}
    I_k \equiv \frac{1}{\sqrt{2\pi}} \int_{-\infty}^{\infty} x^k \e^{-\frac{1}{2}\left(x-2\alpha\right)^2} d x \, .
\end{equation}
Through an integration by parts, a recurrence relation can be obtained for $I_k$. It reads
\begin{equation}
    I_{k+1} = 2\alpha I_k + k I_{k-1}\, ,
\end{equation}
with $I_0(\alpha) = 1$ and $I_1(\alpha) = 2\alpha$. As such, $I_k$ is a $k$-th order polynomial in $\alpha$.
Then, the global minimum of~\eqref{eq:polyn-apdx-lagrange} is such that $\partial \mathcal{L}_N / \partial c_k = 0$ for all $k$, and $\partial \mathcal{L}_N / \partial \lambda = 0$. This corresponds to a linear set of equations in $c_k$ and $\lambda$, and can therefore be rewritten as a problem of the form $A x = y$ where $A$ is a matrix of the $I_k$ integrals, $x = (c_0, \cdots, c_N, \lambda)$ and $y=(0,0,\cdots, 0,-1)$. Such a system is easily solved numerically through matrix inversion, thus yielding the solution of the initial problem.

\end{document}